\documentclass[preprint,3p]{elsarticle}
\biboptions{sort&compress}
\usepackage{listings}
\usepackage{hyperref}
\usepackage{multicol}
\usepackage{lipsum}
\usepackage{breakurl}
\usepackage{multirow, array}
\usepackage{bbm}
\usepackage{graphicx}

\usepackage{xcolor} 

\def\ttus{\char`_}

\definecolor{dkgreen}{rgb}{0,0.6,0}
\definecolor{dred}{rgb}{0.545,0,0}
\definecolor{dblue}{rgb}{0,0,0.545}
\definecolor{lgrey}{rgb}{0.9,0.9,0.9}
\definecolor{gray}{rgb}{0.4,0.4,0.4}
\definecolor{darkblue}{rgb}{0.0,0.0,0.6}
\lstdefinelanguage{cpp}{
      basicstyle=\footnotesize \ttfamily \color{black} \bfseries,   
      breakatwhitespace=false,       
      breaklines=true,               
      captionpos=b,                   
      commentstyle=\color{black},   
      deletekeywords={...},          
      escapeinside={(-@}{@-)},                  
      frame=none, 
      language=C++,                
      keywordstyle=\color{dblue},  
      morekeywords={symbol,numeric,Index,Vector,Symbol,SecDec,ex,string,lst,vector,symtab,std,function,prototype_table,exmap,exvector,XIntegrand,SecDecBase,MinimizeBase,Parser,MapFunction,FeynmanParameter}, 
      identifierstyle=\color{black},
      stringstyle=\color{black},      
      numbers=none,                 
      numbersep=5pt,                  
      numberstyle=\tiny\color{black}, 
      rulecolor=\color{black},        
      showspaces=false,               
      showstringspaces=false,        
      showtabs=false,                
      stepnumber=1,                   
      tabsize=4,                     
    }

\def\cpp{{\tt C++} }

\begin{document}

\begin{frontmatter}

\journal{Computer Physics Communications}

\title{{\tt HepLib}: A {\tt C++} Library for High Energy Physics}

\author[1]{Feng Feng\corref{cor}}
\ead{F.Feng@outlook.com.com}

\author[1]{Yi-Fan Xie}
\author[1]{Qiu-Chen Zhou}
\author[1]{Shan-Rong Tang}

\address[1]{China University of Mining and Technology, Beijing 100083, China}
\cortext[cor]{Corresponding author}

\begin{abstract}
{\tt HepLib} is a \cpp {\tt Lib}rary for computations in {\tt H}igh {\tt E}nergy {\tt P}hysics, it works on top of {\tt GiNaC}, a well-established \cpp library used to perform symbolic computations. {\tt HepLib} combines serval well-known packages to get high efficiency, including {\tt qgraf} to generate Feynman aptitudes, {\tt FORM} to perform Dirac/Color matrix related computations, and {\tt FIRE} or {\tt KIRA} for integration-by-parts (IBP) reduction. Another core feature of {\tt HepLib} lies in the numerical evaluation of master integrals using sector decomposition, which is a general method widely used in high-order numerical computation and has been implemented in many public packages in many different languages, and we present another implementation in the language of \cpp with many new features. We use {\tt GiNaC} to handle the symbolic operations, and export the corresponding integrand into an optimized \cpp code, that will be compiled internally and linked dynamically, a customizable numerical integrator is selected to perform the numerical integration, while the integrand can be evaluated in different float precisions, including the arbitrary precision supported by {\tt MPFR}.
\end{abstract}


\end{frontmatter}

\noindent {\bf Program Summary} \par\vspace{5pt}
\noindent {\it Program title}: {\tt HepLib} \par\vspace{3pt}
\noindent {\it GitHub pages}: {\tt https://heplib.github.io/} \par\vspace{3pt}
\noindent {\it Licensing provisions}: {\tt GPLv3} \par\vspace{3pt}
\noindent {\it Program language}: {\tt C++} \par\vspace{3pt}
\noindent {\it Operating System}: Linux, Mac OS \par\vspace{3pt}
\noindent {\it External libraries}: {\tt GiNaC}, {\tt qgraf}, {\tt MPFR}, {\tt QHull}, {\tt MinUit2}, {\tt CUBA} \par\vspace{3pt}
\noindent {\it External programs:} {\tt FORM}, {\tt Fermat}, {\tt FIRE}, {\tt KIRA}
\par\vspace{3pt}
\noindent {\it Nature of the problem}: There are many independent programs or packages in high energy physics, which are written or developed in different programming languages, here we are trying to provide an integrated interface in the \cpp language, for generating Feynman diagrams/amplitudes, performing algebraic simplifications on Dirac/Color matrix objects, and reducing scalar integrals to master integrals. The evaluation of master integrals is in the core part during the high-order calculations, one needs to resolve ultraviolet/infrared divergence within dimensional regularization and the singularities inside the integration domain while performing multiple dimensional numerical integrations. \par\vspace{3pt}

\noindent {\it Method of solution}: {\tt HepLib} uses {\tt qgraf} to generate the Feynman diagrams/aptitudes, {\tt FORM} to perform Dirac/Color matrix related computations, {\tt FIRE} or {\tt KIRA} for the IBP reduction. A \cpp implementation of sector decomposition method is used to extract both ultraviolet and infrared singularity of the integral, the contour deformation is adopted to avoid the singularity inside the integration domain, a parallelized numerical integration is chosen to achieve high efficiency or performance.\par\vspace{3pt}

\noindent {\it Restrictions}: Depending on the complexity of the problem itself.

\newpage

\section{Introduction}
The higher-order computation in perturbative quantum field theory or Standard Model is indispensable, as the precision or accuracy increases at the Large Hadron Collider, many scattering processes have been calculated up to next-to-next-to-leading order to meet the precision demand, some processes to even higher order~\cite{Amoroso:2020lgh}.
The traditional basic procedures in the higher-order computations in high energy physics involve generating Feynman diagrams/amplitudes, performing algebraic simplifications or traces on Dirac/Color matrix objects, applying partial fraction decomposition and integration-by-parts (IBP) reduction on the scalar integrals to result in the master integrals, finally one needs a way to evaluate those master integrals analytically or numerically.
There are many public packages available to complete those procedures, here we just mention some (not complete) commonly used packages in each step. {\tt FeynArts}~\cite{Hahn:2000kx,FeynArts_website} is a {\tt Mathematica} package used to generate Feynman diagrams/amplitudes, {\tt qgraf}~\cite{Nogueira:1991ex,qgraf_website} is another widely used program in amplitudes generation, which is written in {\tt Fortran} language to get much higher efficiency. {\tt FeynCalc}~\cite{Mertig:1990an,Shtabovenko:2016sxi,FeynCalc_website} developed in {\tt Mathematica} language is commonly used to perform Dirac/Color matrix trace or simplifications, {\tt FORM}~\cite{Vermaseren:2000nd,FORM_website} is another highly efficient program developed in {\tt C} language to handle the operations involving Dirac-$\gamma$ or indexed objects, {\tt FeynCalc}/{\tt FormLink}~\cite{Feng:2012tk,FormLink_website} is written as an interface between {\tt FeynCalc} and {\tt FORM}. The {\tt Mathematica} package {\tt Apart}~\cite{Feng:2012iq,APart_website} can be used to perform partial fractions decomposition. One can use {\tt AIR}~\cite{Anastasiou:2004vj,AIR_website} written in {\tt Maple} language, {\tt FIRE}~\cite{Smirnov:2013dia,Smirnov:2014hma,Smirnov:2019qkx,FIRE_website} written in {\tt Mathematica} and \cpp language, {\tt LiteRed}~\cite{Lee:2012cn,Lee:2013mka,LiteRed_website} a powerful {\tt Mathematica} package, {\tt Reduze}~\cite{Studerus:2009ye,vonManteuffel:2012np,Reduze_website} or {\tt KIRA}~\cite{Maierhoefer:2017hyi,Klappert:2020nbg,KIRA_website}, both in \cpp language, to perform the IBP reductions on scalar integrals to get master integrals. As for the evaluation of master integrals, Mellin-Barnes (MB)~\cite{Smirnov:1999gc,Tausk:1999vh} or sector decomposition (SD)~\cite{Binoth:2003ak,Heinrich:2008si} methods are widely used in numerical calculations. There are many useful subroutines on the {\tt MBTools} website~\cite{MBtools_website} to facilitate the usage of MB method, one can also consult the {\tt AMBRE}~\cite{Dubovyk:2016ocz,Blumlein:2014maa,AMBRE_website} packages for the automatic Mellin-Barnes representation. {\tt FIESTA}~\cite{Smirnov:2008py,Smirnov:2009pb,Smirnov:2013eza,Smirnov:2015mct,FIESTA_website} and {\tt SecDec}~\cite{Borowka:2012yc,Borowka:2013cma,Borowka:2015mxa,SecDec_website} are two widely used packages developed in {\tt Mathematica} language, both provide implementations of SD method, {\tt pySecDec}~\cite{Borowka:2017idc,SecDec_website} is a new version of the program {\tt SecDec}, it uses {\tt Python} as a framework to organize different parts of the evaluation while the main work is done using {\tt FORM} and {\tt CUBA}~\cite{Hahn:2004fe,Hahn:2014fua,CUBA_website}/{\tt QMC}~\cite{Li:2015foa}, {\tt sector\ttus{}decomposition}~\cite{Bogner:2007cr,sector_decomposition_website} is an implementation in \cpp language with the help of {\tt GiNaC}~\cite{Bauer:2000cp,GiNaC_website}, {\tt CSector}~\cite{Gluza:2010rn,CSector_website} is a Mathematica interface to {\tt sector\ttus{}decomposition}. 
The differential equation (DE)~\cite{Kotikov:1991pm,Remiddi:1997ny} method becomes a powerful tool to evaluate the master integrals, especially when the equations can be transformed into a canonical form or $\epsilon$-form~\cite{Henn:2013pwa,Henn:2014qga}, there are also many public packages to find the possible canonical form automatically: {\tt Fuchsia}~\cite{Gituliar:2017vzm,Fuchsia_website}, {\tt epsilon}~\cite{Prausa:2017ltv,epsilon_website}, {\tt CANONICA}~\cite{Meyer:2017joq,CANONICA_website} and so on. The numerical or semi-numerical DE method with powers series expansions can also be found in \cite{Lee:2017qql,Liu:2017jxz}, and {\tt DESS}~\cite{Lee:2017qql,DESS_website} is a {\tt Mathematica} implementation of such a kind of method near singular points. Last but not least, there are also many useful self-integrated programs for the perturbative calculation of cross sections mainly up to next-to-leading order: {\tt FormCalc}~\cite{Hahn:1998yk,FormCalc_website}, {\tt CompHEP}~\cite{Pukhov:1999gg,Boos:2004kh,CompHEP_website}, {\tt FDC}~\cite{Wang:2004du}, {\tt CalcHEP}~\cite{Belyaev:2012qa,CalcHEP_website}, {\tt GoSam}~\cite{Cullen:2011ac, Cullen:2014yla, GoSam_website}, {\tt DIANA}~\cite{DIANA_website}, {\tt MadGraph5\ttus{}aMC@NLO}~\cite{Alwall:2014hca}, {\tt Herwig++}~\cite{Bahr:2008pv}, {\tt SHERPA}~\cite{Gleisberg:2008ta}, {\tt WHIZARD}~\cite{Kilian:2007gr} and many others, we refer the reader to {\tt hepforge projects}~\cite{hepforge_website} for more related programs.

{\tt Mathematica} is a widely used language for generic symbolic or algebraic computations, {\tt Python} and other languages can also be used to achieve the same purpose, while for the numerical evaluation, {\tt Fortran} or {\tt C++/C} is still most favored. Usually, one needs to jump from one language to another during the high order calculations through an interface, here we want to present the library {\tt HepLib} with those interfaces implemented internally, so one does not need to know the internal conversions between different languages. {\tt HepLib} tries to provide a general {\tt C++} framework or interface involving both symbolic computation and numerical evaluation in the high order calculations. 
The core symbolic operations in \cpp language are built on {\tt GiNaC} (we refer the reader to the official tutorial~\cite{ginac_tutorial} for a basic introduction to {\tt GiNaC}). {\tt HepLib} makes small extensions to {\tt GiNaC} by adding some basic classes or objects which are familiar (especially to the users of {\tt FeynCalc}) in the higher-order computations, it provides a few interfaces to other well-known sophisticated programs including {\tt Fermat}~\cite{Fermat_website} to do fast simplifications on multivariate rational polynomials,  {\tt qgraf} to generate Feynman diagrams and amplitudes, {\tt FORM} to handle Dirac/Color object related operations,  {\tt FIRE} or {\tt KIRA} to perform the IBP reductions. As for numerical evaluation, {\tt HepLib} provides another native {\tt GiNaC} implementation of the sector decomposition method with many new features. We will try to add other methods including DE or MB in the near future versions.

\section{Description of {\tt HepLib}}
\subsection{Extension to {\tt GiNaC}}
{\tt HepLib} works on top of {\tt GiNaC} which is a well-established {\tt C++} library for symbolic calculations, in this section we want to give a short introduction to {\tt GiNaC} and make a few extensions to it, while for a general description of {\tt GiNaC}, we refer the reader to its official tutorial~\cite{ginac_tutorial}. 

{\tt GiNaC::ex} is one of the most important and commonly used class in {\tt GiNaC},  it represents a general symbolic expression which contains variables, numbers, functions and so on. The {\tt GiNaC} expressions work like handles, {\it i.e.}, a pointer to the internal actual object, but one needs not to care about the memory allocation and deallocation, since {\tt GiNaC} uses an {\it intrusive reference-counting} mechanism to handle the garbage collection, {\it i.e.}, when an instance of {\tt GiNaC::ex} runs out of its scope, the destructor will check whether any other expression points to such internal actual object too, and deletes the object from memory if that turns out not to be the case. Due to such a reference-counting mechanism, {\tt GiNaC} expressions are not thread-safe, and it would be problematic if one tries to use {\tt GiNaC} from different threads simultaneously.

A {\tt GiNaC::ex} object can contain complex variables instanced from {\tt GiNaC::symbol} class, numbers instanced from {\tt GiNaC::numeric} class, functions instanced from {\tt GiNaC::function} class and so on. Expressions can also be put together to form a new expression or passed as arguments to a function, and so on. Here is a little collection of {\tt GiNaC} expressions (we  omit the namespace {\tt GiNaC} for each class for simplicity):
\begin{lstlisting}[language=cpp]
	symbol x("x"), y("y"), z("z");	//complex variables
    numeric r(2,3); //exact fraction 2/3
	ex e1 = r*x+2*y+pow(y,10); //polynomial in x and y
	ex e2 = (x+1)/(x-1); //rational expression
	ex e3 = sin(x+2*y)+3*z+41; //containing a function
	ex e4 = e3+e2/exp(e1); //expression formed from other expressions
\end{lstlisting}
where the {\tt pow(y,10)} refers to $y^{10}$, and the function {\tt sin} and {\tt exp} are objects of {\tt GiNaC::function}, they are actually {\it pseudo-functions}, {\it i.e.}, their evaluations may be halted if the arguments are not numerical. There are many other mathematical functions defined in {\tt GiNaC}, we refer the reader to the tutorial~\cite{ginac_tutorial} for a complete list.

It should be noted that {\tt GiNaC::symbol} refers to a complex variable, and the {\tt string name} in its constructor {\tt symbol::symbol(string name)} is not the identification to the object, so if we write an expression as {\tt ex e = symbol("x")-symbol("x")}, the two {\tt symbol("x")} are distinct in this expression, the expression {\tt e} is not zero, and one will get something like {\tt x-x} in the output. Furthermore, most of the variables encountered in a physical process are real, so we introduce a class {\tt HepLib::Symbol}, its usage is almost the same as {\tt GiNaC::symbol}, except that it represents a real variable, and the {\tt name} in its constructor {\tt Symbol::Symbol(string name)} is indeed the identification to the object, {\it i.e.}, the expression {\tt Symbol("x")-Symbol("x")} will produce the result {\tt 0} as expected. Finally, we use the following segment of the code {\tt 0.cpp}\footnote{The file {\tt xxx.cpp} in the current version can be found at {\tt codes/xxx.cpp} in {\tt HepLib} archive.} to show the difference of {\tt GiNaC::symbol} and {\tt HepLib::Symbol} in calling the {\tt GiNaC::conjugate} function which returns the complex conjugation of an expression (we also omit the namespace {\tt HepLib} for the corresponding class for simplicity).
 \begin{lstlisting}[language=cpp]
	symbol x("x");	//a complex variable
	Symbol y("y"); //a real variable
    ex e1 = conjugate(x); //complex conjugation of x --> conjugate(x) 
	ex e2 = conjugate(y); //complex conjugation of y --> y
\end{lstlisting}
where one can see {\tt conjugate(y)} will be evaluated to {\tt y}, while {\tt conjugate(x)} will still return as {\tt conjugate(x)}.

{\tt HepLib::Symbol} class is very convenient when one converts a string text into a {\tt GiNaC} expression, and such a conversion usually happens in the data exchange between {\tt HepLib} and other packages ({\it e.g.}, {\tt Fermat} and {\tt FORM}). The conversion from a string text to a {\tt GiNaC} or {\tt HepLib} expression is encoded in the class {\tt HepLib::Parser} which has been introduced as follows:
\begin{lstlisting}[language=cpp]
	class Parser {
	public:
		Parser(); Parser(symtab st); symtab STable;
		ex Read(string instr,bool s2S=true); ex ReadFile(string filename,bool s2S=true);
	};
\end{lstlisting}
where {\tt symtab} is defined in {\tt GiNaC} itself as {\tt typedef std::map<string,ex> symtab}, so the member {\tt STable} is used to map a variable appearing in the string text to the corresponding {\tt ex} object, and an {\it unknown} variable (one does not provide the mapping explicitly in {\tt STable}) will be converted to a {\tt symbol} or {\tt Symbol} object. The method {\tt Read} or {\tt ReadFile} will parse the expression from a string text or a file content respectively, and when {\tt s2S} is {\tt true}, an {\it unknown} variable is converted to a {\tt Symbol} object, which is also the default case. Here is a snip of the code {\tt 1.cpp} to exemplify the usage of {\tt Parser},

\begin{lstlisting}[language=cpp]
	Parser parser;
	string expr_str = "WF(1)+x(1)^2+sin(5)+pow(a,n)";
	ex e1 = parser.Read(expr_str);
	Symbol a("a"), n("n");
	ex e2 = WF(1)+pow(x(1),2)+sin(5)+pow(a,n);
	cout << e1-e2 << endl; // --> 0
\end{lstlisting}
{\tt e1} is now a {\tt GiNaC} expression {\tt x(1)\^{}2+WF(1)+a\^{}n+sin(5)}, the {\it unknown} variables {\tt a} and {\tt n} are parsed to {\tt Symbol("a")} and {\tt Symbol("n")} respectively, and the functions {\tt sin}, {\tt WF} and {\tt x} are also parsed correctly, since {\tt sin} is defined in {\tt GiNaC}, {\tt WF} and {\tt x} are {\it pseudo-functions} introduced in {\tt HepLib}, the both functions can take any {\tt ex} object as the argument and will do nothing on it. We also construct the expression {\tt e2} explicitly to see {\tt e1} and {\tt e2} are indeed identical.

{\tt GiNaC::lst} is another commonly used class that serves for holding a list of arbitrary expressions. One can use the {\tt nops} method to get the number of expressions in a list, the {\tt op} method to access individual elements, and {\tt let\ttus{}op} method to change an element in the list. The {\tt op} and {\tt nops} methods are actually available to any general {\tt GiNaC} expression too, and one can use them to access the internal subexpressions. Here is a snip of the code {\tt 2.cpp} to demonstrate the basic usage of those methods:
\begin{lstlisting}[language=cpp]
	Symbol x("x"), y("y"), z("z"), n("n");
	lst l1 = lst{ x, y, x+z };
	ex e1 = pow(sin(x),n);
	int tot = l1.nops(); // tot is 3
	ex item1 = l1.op(0); // 1st item in l1, --> x
	l1.let_op(2) = e1; // now l1 is {x, y, sin(x)^n}
	tot = e1.nops(); // now tot is 2
	ex item2 = e1.op(1); // 2nd item in e1, --> n
\end{lstlisting}

Note that one can not modify an item in a general expression by using the {\tt let\ttus{}op} method, to achieve this purpose one can use one of the following two methods, one is the expression substitution to make some simple replacements with the help of {\tt GiNaC} function {\tt GiNaC::subs}, the other is to use the class {\tt HepLib::MapFunction} to perform more complicated operations on parts of the expression. For the first method, one usually use the class {\tt GiNaC::wildcard} to do the {\it pattern matching}, a {\tt wildcard} is a special kind of {\tt GiNaC} object which represents an arbitrary expression, and to match different expressions one can use different {\tt label} in its constructor {\tt ex wildcard::wildcard(unsigned label=0)}, one can also use {\tt w} and {\tt w0} in {\tt HepLib} for the abbreviation of {\tt wildcard(0)} object, {\tt w1} for {\tt wildcard(1)}, and so are the {\tt w2} up to {\tt w9}. While for the second method, the {\tt HepLib::MapFunction} class is introduced as follows:
\begin{lstlisting}[language=cpp]
	class MapFunction {
    public:
        ex operator()(const ex &e);
        MapFunction(std::function<ex(const ex &, MapFunction &)>);
    };
\end{lstlisting}
We exemplify both methods by making two substitutions on a simple polynomial $x^4+x^3+x^2+x$, one substitution is to replace any term which {\it explicitly} matches the pattern $x^n$ by $y^{n+2}$, {\it i.e.}, $x^{n} \to y^{2+n}$, the other one is to do the replacement only for the terms with an even exponent, {\it i.e.}, $x^{2n} \to y^{2+2n}$, using the following snip of the code {\tt 3.cpp}:
\begin{lstlisting}[language=cpp]
	Symbol x("x"), y("y");
	//HepLib::str2ex: convert string to expression, using Parser internally
	ex e0 = str2ex("x^4+x^3+x^2+x");
	ex e1 = e0.subs( lst{ pow(x,w)==pow(y,w+2) } ); //(-@ e1 is $y^6+y^5+y^4+x$@-)
	MapFunction map([&](const ex &e, MapFunction self)->ex{ 
		if(e.match(pow(x,w)) && e.op(1).info(info_flags::even)) return pow(y,e.op(1)+2);
		else return e.map(self);
	});
	ex e2 = map(expr); //(-@ e2 is $y^6+x^3+y^4+x$@-)
\end{lstlisting}
where we have used the {\tt C++} {\it lambda} expression as the {\tt std::function} object, and note that the statement {\tt e.map(self)} will apply the specified function on all subexpressions (in the sense of {\tt op()}), non-recursively~\cite{ginac_tutorial}.

Last but not least, one usually gets a long list of expressions ({\it e.g.}, hundreds or thousands of Feynman amplitudes) in high order computations, and performs same or similar operators on each item, since the operations on each item are independent, one can do those operations simultaneously, {\it i.e.}, a single operation executes simultaneously on multiple elements of data (SIMD). As we have mentioned before, one can not use {\tt GiNaC} from different threads simultaneously due to the internal reference-counting mechanism, so one can not run the operations from different threads in one process simultaneously, and we need to turn to a process-based parallel method. We extend {\tt GiNaC} to a parallel version by using the {\tt fork()} system call, {\tt fork()} will create a new process (child process) by duplicating the calling process (parent process), so the input data to the child process is an exact duplicate of the parent process, while we still need a way to get data back from each child process to the parent process. 

{\tt GiNaC} allows creating {\it archives} of expressions which can be stored to or retrieved from files, we wrap these operations into {\tt HepLib} functions {\tt garWrite} (to write an expression into an archive file) and {\tt garRead} (to read an expression from an archive file), for example, {\tt garWrite("e.gar", e)} will export the expression {\tt e} into the archive file {\tt e.gar}, and the file {\tt e.gar} contains all information which is needed to reconstruct the expressions {\tt e}, and {\tt garRead("e.gar")} will return the expression identical to the original {\tt e}. So to get data back from a child process to the parent process, the child process exports the data to an intermediate archive file by using {\tt garWrite}, and the parent process can get the data by using {\tt garRead}. All those steps are encoded in the {\tt HepLib} function  {\tt GiNaC\char`_Parallel} which is introduced as follows:
\begin{lstlisting}[language=cpp]
	vector<ex> GiNaC_Parallel(int ntotal, std::function<ex(int)> f, 
		const string & key="", bool rm=true, const string &pre="  ");
\end{lstlisting}
where {\tt ntotal} is the number of items, {\tt f} is a function taking an {\tt int} as input and {\tt ex} in return, the function {\tt f} will be called on each {\tt i} ({\tt 0}$\leq${\tt i}$<${\tt ntotal}) in parallel, {\tt key} is used to name the directory where the intermediate files store (the process id will be used if {\tt key} is empty, which is also the default value), {\tt rm=true} is to remove the intermediate files at the end of the function call, and the last parameter {\tt pre} refers to some indent space appended to the output message (when the global variable {\tt Verbose} is large than 1). {\tt GiNaC\char`_Parallel} is equivalent to a parallel version of the following {\tt for} statement:
\begin{lstlisting}[language=cpp]
	vector<ex> ret;
	for(int i=0; i<ntotal; i++) ret.push_back(f(i));
	return ret;
\end{lstlisting}
note that the number of parallel processes can be changed by setting the variable {\tt GiNaC\ttus{}Parallel\ttus{}Process}, when {\tt GiNaC\ttus{}Parallel\ttus{}Process} is negative, the number of parallel processes will be set to the number of cpu cores minus 1, which is also the default case, and when {\tt GiNaC\ttus{}Parallel\ttus{}Process} is 0, {\tt GiNaC\char`_Parallel} is equivalent to the {\tt for} statement above. Finally, we end this section by a snip of the code {\tt 4.cpp} to illustrate the basic usage of {\tt GiNaC\ttus{}Parallel}:
\begin{lstlisting}[language=cpp]
	Symbol x("x"), y("y");
    vector<ex> in_vec;
    int total = 1000;
    for(int i=0; i<total; i++) in_vec.push_back(sin(exp(x+y*i)));
    Verbose = 100;
    vector<ex> ret = GiNaC_Parallel(total, [&](int idx)->ex {
        ex data = in_vec[idx];
        data = series_ex(data,x,5); // Taylor expansion around x=0
        return data;
    });
\end{lstlisting}
where we have filled the {\tt in\ttus{}vec} with some simple expressions, and taken the {\it Taylor} expansion on each item parallelly by using {\tt HepLib::ex mma\ttus{}series(const ex \& e, const symbol \&s, int n)} which returns the {\it Taylor} or {\it Laurent} series expansion of {\tt e} up to the {\tt n}-th order with respect to {\tt s} at the point {\tt s=0}.

Before ending this section, it should be noted that the terms order in the output may be different in each run, as stated in {\tt GiNaC} official tutorial~\cite{ginac_tutorial}, for example, the code {\tt cout << x-y;} can produce the output {\tt x-y} or {\tt -y+x}, and to get a {\it fixed} or {\it term-ordered} output, one needs to implement one's own output format, and the class {\tt HepLib::HepFormat} is introduced for such a purpose. The global object {\tt hout} is instanced from {\tt HepFormat} class, its usage is very similar to the standard \cpp object {\tt cout}, and one can use the code {\tt hout << x-y;} to get the {\it fixed} or {\it term-ordered} output.

\subsection{Basic objects in {\tt HepLib}}

\begin{table}[thb!]
\begin{center}
\begin{tabular}{|>{\centering}p{0.15\textwidth}>{\raggedright}p{0.25\textwidth}>{\arraybackslash}p{0.52\textwidth}|}
\hline
\textbf{Object} & \textbf{Example} & \textbf{Description}\\
\hline
{\tt Symbol} & {\tt Symbol("s")} & a real variable {\tt s}.\\
\multirow{3}{*}{\tt Index}
& {\tt Index("mu",Type::VD)} & a Lorentz index {\tt mu} with dimension {\tt D}.\\
& {\tt Index("a",Type::CA)} & a color index {\tt a} with dimension {\tt NA}.\\
& {\tt Index("i",Type::CF)} & a color index {\tt i} with dimension {\tt NF}.\\
{\tt Vector} & {\tt Vector("p")} & a vector/momentum {\tt p}.\\
\multirow{3}{*}{\tt Pair} 
& {\tt Pair(mu,nu)} & a Kronecker delta $\delta_{\mu\nu}$ with {\tt Index} {\tt mu} and {\tt nu}.\\
& {\tt Pair(p,mu)} & a {\tt Vector} {\tt p} with Lorentz {\tt Index} {\tt mu}, {\it i.e.}, ${p}^{\mu}$.\\
& {\tt Pair(p,q)} & a scalar product $p\cdot q$ between {\tt Vector} {\tt p} and {\tt q}.\\
\multirow{2}{*}{\tt SUNT}
& {\tt SUNT(a,i,j)} & a $T$-matrix element $T^{a}_{ij}$ for {\tt SU(N)} group.\\
& {\tt SUNT(lst\{a,b,c\},i,j)} & a matrix element of a product of $T$, {\it i.e.}, $(T^{a}T^{b}T^{c})_{ij}$ \\
{\tt SUNF} & {\tt SUNF(a,b,c)} & a structure constant $f^{abc}$ of {\tt SU(N)} group.\\
{\tt SUNF4} & {\tt SUNF4(a,b,c,d)} & a contract of two {\tt SUNF}, {\it i.e.}, $f^{abe}f^{ecd}$.\\
\multirow{3}{*}{\tt Eps}
& {\tt Eps(mu1,mu2,mu3,mu4)} & a Levi-Civita tensor $\varepsilon_{\mu_1\mu_2\mu_3\mu_4}$. \\
& {\tt Eps(p1,p2,mu1,mu2)} & a partially contracted Levi-Civita tensor $\varepsilon_{p_1p_2\mu_1\mu_2}$.\\
& {\tt Eps(p1,p2,p3,p4)} & a fully contracted Levi-Civita tensor $\varepsilon_{p_1p_2p_3p_4}$.\\
\multirow{3}{*}{\tt DGamma}
& {\tt DGamma(mu,l)} & a Dirac-$\gamma$ matrix $\gamma_{\mu}$ for a fermion line labeled by {\tt l}.\\
& {\tt DGamma(p,l)} & a Dirac slash ${p}\!\!\!/={p^\mu}\gamma_\mu$ for a fermion line labeled by {\tt l}.\\
& {\tt DGamma(1|5|6|7,l)} & a unit matrix, $\gamma_5$,$\gamma_6$,$\gamma_7$ for a fermion line labeled by {\tt l}.\\
\hline
\multirow{3}{*}{\tt SP} 
& {\tt SP(mu,nu)} & evaluated to $\delta_{\mu\nu}$.\\
& {\tt SP(p+s*q,mu)} & evaluated to ${p}^{\mu}+{s}{q}^{\mu}$. \\
& {\tt SP(2*p+q,p+s*q)} & evaluated to $2{p}^2+(2{s}+1){p}\cdot{q}+{s}{q}^2$.\\
\multirow{3}{*}{\tt GAS}
& {\tt GAS(mu)} & evaluated to $\gamma_{\mu}$.\\
& {\tt GAS(3*p+s*q)} & evaluated to $3\,{p}\!\!\!/+{s}{q}\!\!\!/$.\\
& {\tt GAS(1|5|6|7)} & evaluated to a unit matrix, $\gamma_5$,$\gamma_6$,$\gamma_7$, respectively.\\
{\tt LC} & {\tt LC(p,mu,p+s*q,k)} & evaluated to ${s}\varepsilon_{kpq\mu}$ \\
\hline
{\tt TR} & {\tt TR(expr) } & a wrapper for the Dirac trace of expression {\tt expr}.\\
{\tt TTR} & {\tt TTR(lst\{a,b,c,d\})} & a wrapper for the {\tt SU(N)} trace of $T^{a}T^{b}T^{c}T^{d}$. \\
\hline
{\tt form} & {\tt form(expr)} & evaluate the expression {\tt expr} using {\tt FORM} program. \\
\hline
\end{tabular}
\end{center}
\caption{The basic objects defined in {\tt HepLib}. Here, {\tt mu}, {\tt nu} and {\tt mu1} up to {\tt mu4} are objects of Lorentz {\tt Index} class with {\tt Type::VD}, {\tt a}, {\tt b}, {\tt c}, {\tt d} and {\tt e} are objects of {\tt Index} with {\tt Type::CA}, {\tt i} and {\tt j} are objects of {\tt Index} with {\tt Type::CF}, {\tt p}, {\tt q}, {\tt k} and {\tt p1} up to {\tt p4} are objects of {\tt Vector}, and {\tt s} is an object of {\tt Symbol}.\label{heplib_objects}}
\end{table}

The basic objects introduced in {\tt HepLib} (under namespace {\tt HepLib}) for high order computations of high energy physics are listed in Table~\ref{heplib_objects}, including the following basic classes: {\tt Index} (a Lorentz index or color index of {\tt SU(N)}), {\tt Vector} (a vector or momentum), {\tt Pair} (a metric tensor or scalar product between vectors), {\tt SUNT} (a {\tt T} matrix element of {\tt SU(N)}), {\tt SUNF} (a structure constant of {\tt SU(N)}), {\tt Eps} (a Levi-Civita tensor) and {\tt DGamma} (a Dirac-$\gamma$ matrix in a specific fermion line). To facilitate the usage of these classes, the functions {\tt SP}/{\tt GAS}/{\tt LC} can be used to construct the {\tt Pair}/{\tt DGamma}/{\tt Eps} objects directly from the expressions involving {\tt Index} or {\tt Vector} objects respectively, the linear expressions provided to the arguments of those functions will be expanded automatically. The {\it pseudo-functions} {\tt TR} and {\tt TTR} are provided as function wrappers for the Dirac trace and {\tt SU(N)} trace respectively, and all the actual evaluations (both trace and contraction) will be performed by {\tt FORM} program internally through the interface {\tt form}. These objects introduced above are similar to the ones in {\tt FORM} program and {\tt FeynCalc} package with the correspondence in Table~\ref{heplib_form_fc}.

\begin{table}[thb!]
\begin{center}
\begin{tabular}{|c|ccccccc|}
\hline
{\bf\tt HepLib} & {\tt Index} & {\tt Vector} & {\tt Pair} & {\tt Eps} & {\tt SUNT} & {\tt SUNF} & {\tt DGamma} \\
{\tt FORM} & {\tt Index} & {\tt Vector} & {\tt d\ttus} & {\tt e\ttus} & {\tt T} & {\tt f}  & {\tt g\ttus} \\
{\bf\tt FeynCalc} & {\tt LorentzIndex} & {\tt Momentum} & {\tt MTD/FVD/SPD} & {\tt LC} & {\tt SUNT} & {\tt SUNF} & {\tt DiracGamma} \\
\hline
\end{tabular}
\end{center}
\caption{The correspondence between {\tt HepLib} and {\tt FORM} or {\tt FeynCalc}.\label{heplib_form_fc}}
\end{table}

There are a few things that need to be noted:
\begin{itemize}
\item Similar to the {\tt Symbol} class introduced in the previous section, two objects of class {\tt Index}, {\tt Vector}, {\tt SUNT} and so on are equal or identical if the inputs provided to the constructor are equal, {\it e.g.}, if we defined two {\tt Vector} objects with {\tt Vector p1("p"), p2("p");}, then {\tt p1} and {\tt p2} are identical. 
\item There is no need to introduce an extra operator for the product of two {\it non-commutative} {\tt DGamma} objects, the operator {\tt *} between {\tt DGamma} objects will be treated as {\it non-commutative} multiplication automatically. 
\item Sometimes, it may be useful to use the {\it commutative} matrix elements, instead of the {\it non-commutative} matrix itself, for example in generating Feynman amplitudes, the {\it pseudo-functions} {\tt Matrix(M,i,j)} (the matrix element $M_{ij}$) has been introduced for this purpose, the multiplication of {\tt Matrix} objects is {\it commutative}, and {\tt MatrixContract} can contract the matrix indices and result in the properly ordered {\it non-commutative} objects. 
\item All actual operations involving Lorentz index contraction or trace of Dirac/Color matrix will be handled by {\tt FORM} program internally through the function call {\tt form(expr)}. The internal details behind the {\tt form} function is almost the same as {\tt FormLink}: two {\tt pipes} will be created between {\tt HepLib} and {\tt FORM} program for data exchange, the input {\tt expr} will be translated into a {\tt FORM} script and sent to {\tt FORM} for execution through one of the {\tt pipes}, and the output will be transferred back to {\tt HepLib} through the other {\tt pipe} and convert back to {\tt HepLib} objects at the finish of {\tt FORM} execution. To avoid the expanse in the creation and destruction of processes again and again, we keep the {\tt FORM} process alive after the first call to {\tt form} during the entire program life. Since the {\tt FORM} program will be invoked internally, we need the shell command {\tt form} can be found in the environment variable {\tt PATH} in order to use {\tt form} function from {\tt HepLib}.

\item The default dimension for a Lorentz {\tt Index} is {\tt D} (represented by {\tt Symbol("D")}), and the default conversion from {\tt TR} to {\tt FORM} is {\tt tracen}, so an {\tt Error} will be thrown if one uses $\gamma_5$ in {\tt TR}, while one can change the default dimension to {\tt 4} by the statement {\tt form\ttus{}using\ttus{}dim4=true;} before calling the function {\tt form}, and the {\tt trace4} will be used for the Dirac trace in that case. Furthermore, {\tt HepLib} does not distinguish the upper and lower index objects, when the contraction occurs between two {\tt Index} objects, one index is supposed to be an upper index and the other is supposed to be a lower index.

\item The {\tt FORM} procedure {\tt SUn.prc} from the {\tt color} package~\cite{color_package,color_website,Cvitanovic:1976am} has been used for the treatment of {\tt SUNT} and {\tt SUNF} color objects.

\item The function {\tt conjugate} is introduced in {\tt GiNaC} itself, this function has been overridden for the objects introduced in Table~\ref{heplib_objects}, it can be used for complex conjugation involving Dirac-$\gamma$ and {\tt SUNT} objects, it is similar to the function {\tt ComplexConjugate} in {\tt FeynCalc}.

\item To facilitate the conversion to {\tt FeynCalc}, the {\tt FCFormat} class is provided in {\tt HepLib}, the global object {\tt fcout} is an instance of {\tt FCFormat}, which is similar to {\tt hout} or the standard \cpp {\tt cout} object, one can use {\tt fcout} to print an expression {\tt expr} in {\tt FeynCalc} format with {\tt fcout << expr << endl;}.

\end{itemize}

We end this section by a physical expression: the squared matrix element $\vert{\cal M}\vert^2$ for the  process $e^-(p) + e^+(p^\prime) \to \gamma^*(q) \to \mu^+(k) + \mu^-({k^\prime})$, which is taken from {\tt (5.4)} on page {\tt 132} of the textbook of quantum field theory~\cite{Peskin_book}:
\begin{equation}
\frac{1}{4}\sum_{\rm spins}\vert{\cal M}\vert^2 = \frac{e^4}{4q^4} {\rm tr}\Big[ (p\!\!\!/^\prime-m_e)\gamma^\mu(p\!\!\!/+m_e)\gamma^\nu \Big]\, {\rm tr}\Big[ (k\!\!\!/+m_\mu)\gamma_\mu(k\!\!\!/^\prime-m_\mu)\gamma_\nu \Big]
\label{squared_matrix_element}
\end{equation}
We use the following snip of the code {\tt 5.cpp} to illustrate how to input the expression above in {\tt HepLib}:
\begin{lstlisting}[language=cpp]
    Symbol me("me"), mm("mm"), e("e");
	Index mu("mu"), nu("nu");
	Vector p("p"), P("P"), k("k"), K("K"), q("q");
	letSP(p)=me*me; letSP(P)=me*me;
	letSP(k)=mm*mm; letSP(K)=mm*mm;
	#define gpm(p,m) (GAS(p)+m*GAS(1))
	ex tr1 = TR( gpm(P,-me)*GAS(mu)*gpm(p,me)*GAS(nu) );
	ex tr2 = TR( gpm(k,mm)*GAS(mu)*gpm(K,-mm)*GAS(nu) );
	ex res =  pow(e,4) / (4*pow(SP(q),2)) * tr1 * tr2;
    form_using_dim4 = true; //using 4 dimension
	res = form(res); //using form(res,n) with n>1 to print FORM script
    res = exfactor(res);
    hout << res.subs(me==0) << endl; // note hout is used
\end{lstlisting}
where the {\tt Vector} objects {\tt p}, {\tt P}, {\tt k}, {\tt K} and {\tt q} refer to the momenta $p$, $p^\prime$, $k$, $k^\prime$ and $q$ respectively, the {\tt Index} objects {\tt mu} and {\tt nu} refer to the Lorentz indices $\mu$ and $\nu$ respectively, the symbols {\tt me}, {\tt mm} and {\tt e} refer to $m_e$, $m_\mu$ and $e$ respectively. We also use {\tt letSP} to assign the scalar product between momenta (to clear the {\tt SP} assignment, one can use {\tt clearSP}). After the function call to {\tt form}, the expression will be translated to {\tt FORM} script (to view the {\tt FORM} script, one can replace the function call {\tt form(res)} with {\tt form(res,10)}) and evaluated by {\tt FORM} program, and finally one gets the output as {\tt 8*e\^{}4*q.q\^{}(-2)*(k.P*p.K+mm\^{}2*p.P+P.K*p.k)}, which corresponds to the final result in {\tt (5.10)} on page {\tt 135} of textbook~\cite{Peskin_book}.
\begin{equation}
\frac{1}{4}\sum_{\rm spins}\vert{\cal M}\vert^2 = \frac{8e^4}{q^4} \Big[ (p\cdot k)(p^\prime\cdot k^\prime) + (p\cdot k^\prime)(p^\prime\cdot k)+m_\mu^2 (p\cdot p^\prime) \Big]
\end{equation}
One needs the {\tt SU(N)} group factor ${\rm tr}(t^a t^b)$ when one replaces $\mu^+\mu^-$ by $q\bar{q}$, {\it i.e.}, the process $e^+e^-\to q\bar{q}$ (see {\tt (17.6)} on page {\tt 549} of \cite{Peskin_book}), and one can use the {\it pseudo-function} {\tt TTR} for the trace of a product of {\tt SU(N)} {\tt T}-matrix objects:
\begin{lstlisting}[language=cpp]
    Index a("a",Index::Type::CA);
 	ex tr = TTR(lst{a,a});
	tr = form(tr);
\end{lstlisting}
Another way to achieve the same purpose is to use the {\tt SU(N)} {\tt T}-matrix element, {\it i.e.}, ${\rm tr}(t^a t^a) = t^a_{ij} t^a_{ji}$, as follows:
\begin{lstlisting}[language=cpp]
    Index a("a",Index::Type::CA), i("i",Index::Type::CF), j("j",Index::Type::CF);
 	ex tr = SUNT(a,i,j) * SUNT(a,j,i); // or SUNT(lst{a,a},i,i)
	tr = form(tr);
\end{lstlisting}

\subsection{Generating Feynman diagrams/amplitudes}

The namespace {\tt HepLib::QGRAF} is introduced for the generation of Feynman diagrams/amplitudes for the underlying scattering process. The main interface to {\tt qgraf} program is encoded in the class {\tt Process} introduced as follows:
\begin{lstlisting}[language=cpp]
	class Process {
	public:
		string Model; string In; string Out; string LoopPrefix = "q";
		int Loops; string Options; vector<string> Others;
		lst Amplitudes(symtab st, bool debug=false);
	}
\end{lstlisting}
Most of the members are the same as those in the input file {\tt qgraf.dat} of {\tt qgraf} program, here we exemplify the basic usage by the process $e^-(p) + e^+(P) \to \gamma^*(q) \to \mu^+(k) + \mu^-(K)$ introduced in the previous section, using the following snip of the code {\tt 6.cpp}:
\begin{lstlisting}[language=cpp]
	Vector p("p"), P("P"), k("k"), K("K");
    Process proc;
    proc.Model = R"EOF(
        [e, ebar, -]
        [mu, mubar, -]
        [A, A, +]
        [ebar, e, A]
        [mubar, mu, A]
        )EOF";
    proc.In = "e[p],ebar[P]";
    proc.Out = "mubar[k],mu[K]";
    proc.Options = "onshell";
    proc.Loops = 0;
	symtab st;
    st["p"] = p; st["P"] = P; st["k"] = k; st["K"] = K;
    auto amps = proc.Amplitudes(st);
\end{lstlisting}
where we define the {\tt QED} model involving three particles {\tt e} ($e^-$) and {\tt mu} ($\mu^-$) and {\tt A} ($\gamma$), and their anti-particles {\tt ebar} ($e^+$) and {\tt mubar} ($\mu^+$) as well, and define the {\tt In} and {\tt Out} states for our process respectively.
The function {\tt Amplitudes} will invoke {\tt qgraf} program internally, and use the following style\footnote{One can modify the style through the static member {\tt Process::Style}, but if one introduces new functions other than {\tt Propagator}, {\tt Vertex} and so on,  one needs to define these new functions as {\tt GiNaC} {\it pseudo-functions}.} to control and generate the output,
\begin{lstlisting}[language=cpp]
	<diagram>
	((<sign><symmetry_factor>)*
	<in_loop>InField(<field>,<field_index>,<momentum>)*
	<end>
	<back><out_loop>OutField(<field>,<field_index>,<momentum>)*
	<end>
	<back><propagator_loop>Propagator(
	<back>Field(<field>,<field_index>),
	<back>Field(<dual-field>,<dual-field_index>),
	<back><momentum>)*
	<end>
	<back><vertex_loop>Vertex(
	<back><ray_loop>Field(<field>,<field_index>,<momentum>),
	<back><end><back>)*
	<end><back><back>)
	,
	<epilogue>
	<exit>
\end{lstlisting}
and finally parse the output to {\tt HepLib} objects wrapped in the {\it pseudo-functions} ({\tt Field}, {\tt Propagator}, {\tt Vertex}, {\it etc.}). Note that one needs to provide a {\tt symtab} object as the argument in calling {\tt Amplitudes}, since we want the objects {\tt p},  {\tt k} and so on to be parsed as {\tt Vector} instead of {\tt Symbol}. All amplitudes are returned as a {\tt lst} ({\tt amps} above), and the typical {\tt amps} looks as follows:
\begin{lstlisting}[language=cpp]
  { -InField(e,-1,p)*InField(ebar,-3,P)*OutField(mu,-4,K)*OutField(mubar,-2,k)*Propagator(Field(A,1),Field(A,2),-P-p)*Vertex(Field(ebar,-3,P),Field(e,-1,p),Field(A,1,-P-p))*Vertex(Field(mubar,-4,-K),Field(mu,-2,-k),Field(A,2,P+p)) }
\end{lstlisting}
where one can see the {\tt field\ttus{}index} appearing in the second argument of function {\tt Field}, and note that {\tt qgraf} program labels the incoming particles as {\tt -1}, {\tt -3}, $\cdots$, and outgoing particles as {\tt -2}, {\tt -4}, $\cdots$, any {\tt field\ttus{}index} associated with an internal line is positive. 

Currently, {\tt HepLib} does not provide other kinds of diagram filtering options than what is available in {\tt qgraf} program itself. One can use the function {\tt TopoLines} to generate the topological lines from an amplitude, and draw the corresponding Feynman diagram with the help of the \LaTeX{} package {\tt Tikz-Feynman}~\cite{Ellis:2016jkw,Tikz_Feynman_website}, all the internals are encoded in the function call {\tt DrawPDF}, as exemplified by the snip of the code {\tt 6.cpp}:
\begin{tabular}{>{\centering}m{0.63\textwidth}>{\centering\arraybackslash}m{0.3\textwidth}}
\begin{lstlisting}[language=cpp]
    InOutTeX[-1]="$e^-(p)$";
    InOutTeX[-3]="$e^+(P)$";
    InOutTeX[-2]="$\\mu^+(k)$";
    InOutTeX[-4]="$\\mu^-(K)$";
    LineTeX[e] = "fermion, edge label=$e$";
    LineTeX[ebar] = "anti fermion, edge label=$e$";
    LineTeX[mu] = "fermion, edge label=$\\mu$";
    LineTeX[mubar] = "anti fermion, edge label=$\\mu$";
    LineTeX[A] = "photon, edge label=$\\gamma$";
    DrawPDF(amp, "amps.pdf");
\end{lstlisting} & \includegraphics[width=0.23\textwidth]{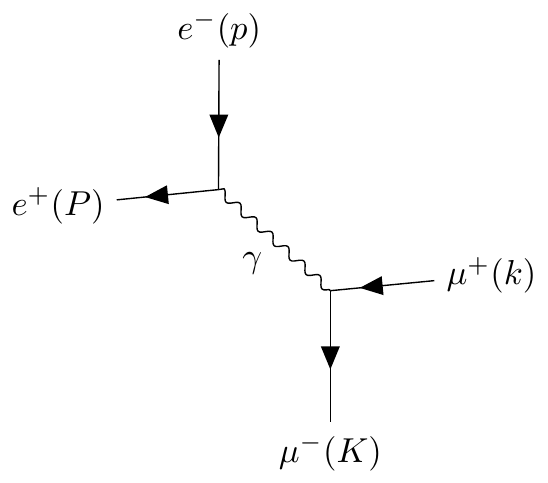}
\end{tabular}\\
note that, since the {\tt Tikz-Feynman} package is used, it is required that the shell command {\tt lualatex} and {\tt pdflatex} should be located in the environment variable {\tt PATH}, and {\tt Tikz-Feynman} package should be installed as well.

As for the returned amplitudes from {\tt Amplitudes}, one still needs to provide the corresponding Feynman rules explicitly, as illustrated in the following snip of the code {\tt 6.cpp}:
\begin{lstlisting}[language=cpp]
	auto amps_feyn_rule = MapFunction([&](const ex &e, MapFunction &self)->ex {
        if(isFunction(e,"OutField") || isFunction(e,"InField")) return 1;
        else if(isFunction(e, "Propagator")) {
            auto fi1 = e.op(0).op(1);
            auto fi2 = e.op(1).op(1);
            auto mom = e.op(2);
            if(e.op(0).op(0)==A) {
                return (-I) * SP(LI(fi1),LI(fi2))/SP(mom); //Feynman Gauge
            } else if(e.op(0).op(0)==ebar) {
                return I * Matrix(GAS(mom)+GAS(1)*me,DI(fi1),DI(fi2))/(SP(mom)-me*me);
            } else if(e.op(0).op(0)==mubar) {
                return I * Matrix(GAS(mom)+GAS(1)*mm,DI(fi1),DI(fi2))/(SP(mom)-mm*mm);
            }
        } else if(isFunction(e, "Vertex")) {
            auto fi1 = e.op(0).op(1);
            auto fi2 = e.op(1).op(1);
            auto fi3 = e.op(2).op(1);
            if(e.op(0).op(0)==ebar) {
                return I*Symbol("e")*Matrix(GAS(LI(fi3)),DI(fi1),DI(fi2));
            } else if(e.op(0).op(0)==mubar) {
                return I*Symbol("e")*Matrix(GAS(LI(fi3)),DI(fi1),DI(fi2));
            }
        } 
		return e.map(self);
    })(amps);
\end{lstlisting}
where the uniqueness of {\tt field\ttus{}index} from {\tt qgraf} program is used to label the Lorentz index, Dirac-$\gamma$ or Color index (through the functions {\tt LI}/{\tt DI}/{\tt CI}\footnote{Each function accepts an integer as input, {\it e.g.}, {\tt LI(1)} refers to {\tt Index("li1")}, {\tt LI(-1)} refers to {\tt Index("lim1")}, more details can be found in {\tt HepLib} document~\cite{HepLib_document}.}), and the final expression of {\tt amps\ttus{}feyn\ttus{}rule} looks like:
\begin{lstlisting}[language=cpp]
 {-I*e^2*Matrix(((-@$\gamma$@-).li1),dim3,dim1)*Matrix(((-@$\gamma$@-).li2),dim4,dim2)*li2.li1*(P.P+p.p+2*p.P)^(-1)}
\end{lstlisting}
and to get the squared matrix element in (\ref{squared_matrix_element}), one still needs to multiply it by its conjugation and the spin polarization summation as follows:
\begin{lstlisting}[language=cpp]
	ex ampL = amps_feyn_rule.op(0);
	ex ampR = IndexL2R(conjugate(ampL));
	#define SS1(p,m,i) Matrix(GAS(p)+m*GAS(1),DI(i),RDI(i))
	#define SS2(p,m,i) Matrix(GAS(p)+m*GAS(1),RDI(i),DI(i))
	ex M2 = ampL*ampR*SS1(p,me,-1)*SS2(P,-me,-3)*SS1(k,-mm,-2)*SS2(K,mm,-4);
	M2 = MatrixContract(M2);
\end{lstlisting}
note that the function {\tt IndexL2R} (usually used in the case that the dummy index needs to be remapped) will change, and only change, the objects generated from {\tt LI/DI/CI} functions to their {\it conjugated} parts, {\it e.g.}, {\tt LI(1)} (equivalent to {\tt Index("li1")}) will be mapped to {\tt RLI(1)} (equivalent to {\tt Index("rli1")}), and so on.

Last but not least in this section, to illustrate the operations on more complicated indexed objects, or IBP reduction for loop corrections in the following sections, let's consider another physical process $\psi \to \gamma^* \to e^+ e^-$~\cite{Czarnecki:1997vz,Beneke:1997jm}. The form factor for this process can be extracted from the amplitude of the corresponding partonic process $Q(p)+\bar{Q}(p) \to \gamma^*(2p)$ ($Q$ refers to the heavy quark), and one can use the following {\tt Process} object to get the related amplitudes up to next-to-next-leading order:\footnote{All code can be found at {\tt codes/nnlo.cpp} in {\tt HepLib} archive.}
\begin{lstlisting}[language=cpp]
	Vector p("p"), q1("q1"), q2("q2");
	ex p1 = p, p2 = p;
	QGRAF::Process proc;
    proc.Model = model;
    proc.In = "Q[p1],Qbar[p2]";
    proc.Out = "A[pA]";
    proc.Options = "onshell";
    proc.LoopPrefix = "q";
    symtab st;  st["pA"] = p1+p2;
    st["p1"] = p1; st["p2"] = p2; st["q1"] = q1; st["q2"] = q2;
    proc.Loops = 0; auto amp0 = proc.Amplitudes(st);
    proc.Loops = 1; auto amp1 = proc.Amplitudes(st);
    proc.Loops = 2; auto amp2 = proc.Amplitudes(st);
\end{lstlisting}
and the underlying {\tt model} is defined as follows:
\begin{lstlisting}[language=cpp]
	std::string model = R"EOF(
		[q, qbar, -]
		[Q, Qbar, -]
		[gh, ghbar, -]
		[g, g, +, notadpole]
		[A, A, +, external]
		[qbar, q, g; QCD='+1']
		[Qbar, Q, g; QCD='+1']
		[g, g, g, g; QCD='+2']
		[g, g, g; QCD='+1']
		[ghbar, gh, g; QCD='+1']
		[qbar, q, A; QCD='+0']
		[Qbar, Q, A; QCD='+0']
	)EOF";
\end{lstlisting}
where {\tt q} refers to the light quark, {\tt Q} to heavy quark, {\tt g} to gluon, and {\tt gh} to the ghost of gluon, furthermore we define the photon field as {\it external} to avoid {\tt QED} loop corrections. One can use the following {\tt MapFunction} object to assign the corresponding Feynman rules to the amplitudes {\tt amp0} (leading order), {\tt amp1} (next-to-leading order) and {\tt amp2} (next-to-next-to-leading order):
\begin{lstlisting}[language=cpp]
auto map = MapFunction([&](const ex &e, MapFunction &self)->ex{
    if(isFunction(e,"OutField")||isFunction(e,"InField")) return 1;
    else if(isFunction(e,"Propagator")){
        if(e.op(0).op(0)==q) return QuarkPropagator(e,0);
        else if(e.op(0).op(0)==Q) return QuarkPropagator(e,m);
        else if(e.op(0).op(0)==g) return GluonPropagator(e);
        else if(e.op(0).op(0)==gh) return GhostPropagator(e);
    }else if(isFunction(e,"Vertex")){
        if(e.nops()==3&&e.op(0).op(0)==ghbar && e.op(1).op(0)==gh){
            return gh2gVertex(e); // ghbar-gh-g
        }else if(e.nops()==3&&e.op(0).op(0)==g&&e.op(1).op(0)==g&&e.op(2).op(0)==g){
            return g3Vertex(e); // g^3
        }else if(e.nops()==4&&e.op(0).op(0)==g&&e.op(1).op(0)==g&&e.op(2).op(0)==g){
            return g4Vertex(e); // g^4
        }else if(e.nops()==3 &&((e.op(0).op(0)==qbar)||(e.op(0).op(0)==Qbar))){
            if(e.op(2).op(0)==g) return q2gVertex(e); // qbar-q-g or Qbar-Q-g
            else{ // qbar-q-A or Qbar-Q-A
                ex fi1 = e.op(0).op(1), fi2 = e.op(1).op(1), fi3 = e.op(2).op(1);
                return Matrix(GAS(LI(fi3)),DI(fi1),DI(fi2)) * SP(TI(fi1),TI(fi2));
            }
        }  
    } 
	return e.map(self);
});
amp0 = ex_to<lst>(map(amp0)); amp1 = ex_to<lst>(map(amp1)); amp2 = ex_to<lst>(map(amp2));
\end{lstlisting}
where the functions {\tt xxxPropagator}/{\tt xxxVertex} under namespace {\tt HepLib::QGRAF} correspond to the Feynman rules for the propagator/vertex in {\tt QCD}, and the first argument for the {\tt xxxPropagator}/{\tt xxxVertex} are always {\tt Propagator}/{\tt Vertex} objects directly generated from {\tt Amplitudes}, we refer the reader to {\tt HepLib} document~\cite{HepLib_document} for detailed information.

Finally, to save the amplitudes for later processing, one can export the whole list to a {\tt gar} file with the function {\tt garWrite}, and use {\tt garRead} to get the data back as follows:
\begin{lstlisting}[language=cpp]
	ex amps = lst { amp0, amp1, amp2 };
	garWrite("amps.gar", amps);
	amps = garRead("amps.gar"); // read back to amps
\end{lstlisting}
\subsection{More operations on indexed objects}

Generally, one gets the amplitude with {\it non-commutative} objects wrapped in the {\it pseudo-function} {\tt Matrix} with {\it matrix} index, and one can use {\tt MatrixContract} to perform the {\it matrix} multiplication, while for the Lorentz or Color indexed objects, the related operations are handled by {\tt FORM} program internally. In this section, we take {\tt amp1} (the amplitude for $Q\bar{Q}\to\gamma^*$ at next-to-leading order) in last section as an example to illustrate more operations on the indexed objects. Note that all the \cpp outputs in this section are generated by the code {\tt codes/nlo.cpp} in {\tt HepLib} archive.

One can use {\tt garRead} to get {\tt amp1} back from the {\tt gar} file as follows:
\begin{lstlisting}[language=cpp]
(-I)*gs^2*Matrix(((-@$\gamma$@-).li1),di4,dim1)*Matrix(((-@$\gamma$@-).li2),dim3,di5)*Matrix(((-@$\gamma$@-).mu),di6,di3)*Matrix(((-@$\gamma$@-).p)+m*(-@$\mathbbm{1}$@-)+(-1)*((-@$\gamma$@-).q1),di3,di4)*Matrix(m*(-@$\mathbbm{1}$@-)+(-1)*((-@$\gamma$@-).p)+(-1)*((-@$\gamma$@-).q1),di5,di6)*ci2.ci1*li2.li1*ti4.ti3*ti6.ti3*ti6.ti5*q1.q1^(-1)*T(ci1,ti4,tim1)*T(ci2,tim3,ti5)*(p.p+q1.q1+(-1)*m^2+(-2)*q1.p)^(-1)*(p.p+q1.q1+(-1)*m^2+2*q1.p)^(-1)
\end{lstlisting}
where {\tt q1} is the loop momentum, and {\tt mu} is the Lorentz index associated with the virtual photon $\gamma^*$, and one can get a more compact expression as follows, with the help of function call {\tt MatrixContract}:
\begin{lstlisting}[language=cpp]
(-I)*gs^2*Matrix(((-@$\gamma$@-).li2)*(m*(-@$\mathbbm{1}$@-)+(-1)*((-@$\gamma$@-).p)+(-1)*((-@$\gamma$@-).q1))*((-@$\gamma$@-).lim2)*(((-@$\gamma$@-).p)+m*(-@$\mathbbm{1}$@-)+(-1)*((-@$\gamma$@-).q1))*((-@$\gamma$@-).li1),dim3,dim1)*ci2.ci1*li2.li1*ti4.ti3*ti6.ti3*ti6.ti5*q1.q1^(-1)*T(ci1,ti4,tim1)*T(ci2,tim3,ti5)*(p.p+q1.q1+(-1)*m^2+(-2)*q1.p)^(-1)*(p.p+q1.q1+(-1)*m^2+2*q1.p)^(-1)
\end{lstlisting}
as one can see there is only one {\tt Matrix} object left with {\it external} field index {\tt dim3} and {\tt dim1}, which needs to be contracted with the Dirac spinors $\bar{v}_{\tt dim3}$ (from heavy anti-quark) and $u_{\tt dim1}$ (from heavy quark) respectively. We use the the spin projector technique~\cite{Bodwin:2013zu,Bodwin:2010fi} to make the following replacement:
$$
u_{\tt dim1} \bar{v}_{\tt dim3} \to \left[ -\frac{1}{4\sqrt{2}m^2}(p\!\!\!/+m)\varepsilon\!\!\!/(p\!\!\!/-m) \right]_{\tt dim1\,dim3} \!\!\!\!\!\! \to -\frac{1}{4\sqrt{2}m^2} \varepsilon^\nu\, {\tt Matrix}((p\!\!\!/+m)\gamma_\nu(p\!\!\!/-m),{\tt dim1},{\tt dim3})
$$
note that one can pull/drop out the overall polarization vector $\varepsilon^\nu$ in the actual computation, and get the result involving the Dirac trace {\tt TR} after using {\tt MatrixContract} again: 
\begin{lstlisting}[language=cpp]
(-I)*NF^(-1/2)*gs^2*TR((1/32)*sqrt(2)*m^(-3)*(((-@$\gamma$@-).li2)*(m*(-@$\mathbbm{1}$@-)+(-1)*((-@$\gamma$@-).p)+(-1)*((-@$\gamma$@-).q1))*((-@$\gamma$@-).mu)*(((-@$\gamma$@-).p)+m*(-@$\mathbbm{1}$@-)+(-1)*((-@$\gamma$@-).q1))*((-@$\gamma$@-).li1)*(((-@$\gamma$@-).p)+m*(-@$\mathbbm{1}$@-))*(2*m*(-@$\mathbbm{1}$@-)+2*((-@$\gamma$@-).p))*((-@$\gamma$@-).nu)*(((-@$\gamma$@-).p)+(-1)*m*(-@$\mathbbm{1}$@-))))*ci2.ci1*li2.li1*ti4.ti3*ti6.ti3*ti6.ti5*tim3.tim1*q1.q1^(-1)*T(ci1,ti4,tim1)*T(ci2,tim3,ti5)*(p.p+q1.q1+(-1)*m^2+(-2)*q1.p)^(-1)*(p.p+q1.q1+(-1)*m^2+2*q1.p)^(-1)
\end{lstlisting}
one can send the expression to {\tt FORM} program through {\tt form} function, and get the evaluated result as follows:
\begin{lstlisting}[language=cpp]
(8/3*I)*sqrt(2)*sqrt(3)*gs^2*q1.q1^(-1)*(q1.q1+2*q1.p)^(-1)*(2*q1.p+(-1)*q1.q1)^(-1)*(2*m^2*nu.mu+2*q1.mu*q1.nu+(-2)*ep*q1.mu*q1.nu+(-1)*nu.mu*q1.q1+ep*nu.mu*q1.q1)
\end{lstlisting}
furthermore, one can write down the expression above in a more readable form:
\begin{equation}
\sqrt{\frac{2}{3}} 8 i g_s^2 \frac{1}{q_1^2} \frac{1}{2p\cdot q_1-q_1^2} \frac{1}{2p\cdot q_1+q_1^2} \left\{ \left[ (\epsilon-1) q_1^2 + 2m^2 \right] g^{\mu\nu} - 2(\epsilon-1) q_1^\mu q_1^\nu \right\} \label{mma_expr}
\end{equation}
with the help of the {\tt Mathematica} package {\tt HepLib.m} using the following {\tt Mathematica} code:
\begin{lstlisting}[language=cpp]
    Get["https://heplib.github.io/HepLib.m"]; "output from HepLib"//C2M//Simplify
\end{lstlisting}
One can see the appearance of tensor integrals (the integrals with free indices) involving $q_1^\mu$ or $q_1^\nu$ in (\ref{mma_expr}), those can not be handled directly by IBP reduction tools, so one needs to transfer the Lorentz indices from loop momenta to external momenta in the loop integration, we call it {\it Tensor Index Reduction} (TIR). We illustrate the basic ideas behind TIR by the tensor integral above, by writing down the following substitution according to the Lorentz invariance:
\begin{equation}
 q_1^\mu q_1^\nu \to c_{00}(p^2,p\cdot q_1,q_1^2) \, g^{\mu\nu}  + c_{11}(p^2,p\cdot q_1,q_1^2) \, p^\mu p^\nu \label{TIR}
\end{equation}
and multiple both sides of (\ref{TIR}) by $g^{\mu\nu}$ and $p^\mu p^\nu$, and get coefficients $c_{00}$ and $c_{11}$ by solving the resulting linear equations. It should be noted that TIR is only valid when we perform loop integration on both sides, {\it i.e.},
\begin{equation}
\int \frac{d^D q_1}{(2\pi)^D} \, {\cal F}(p^2,p\cdot q_1,q_1^2) \, q_1^\mu q_1^\nu = g^{\mu\nu}\int \frac{d^D q_1}{(2\pi)^D} \, c_{00} {\cal F}(p^2,p\cdot q_1,q_1^2) + p^\mu p^\nu\int \frac{d^D q_1}{(2\pi)^D} \, c_{11} {\cal F}(p^2,p\cdot q_1,q_1^2) \nonumber
\end{equation}
where ${\cal F}$ is any function depending on the scalar products $p^2$, $p\cdot q_1$ and $q_1^2$. The {\it Tensor Index Reduction} in {\tt HepLib} is encoded in the function {\tt TIR}\footnote{The related functions in {\tt FeynCalc} are known as {\tt TID} and {\tt TIDL}.} introduced as follows:
\begin{lstlisting}[language=cpp]
	ex TIR(const ex &expr_in, const lst &loop_ps, const lst &ext_ps);
\end{lstlisting}
where {\tt loop\ttus{}ps} refers to the list of loop momenta, {\tt ext\ttus{}ps} to the list of external momenta, and one can get the result as follows, after invoking function call {\tt TIR(amp1,lst\{q1\},lst\{p\})}:
\begin{equation}
\sqrt{\frac{2}{3}} 8 i g_s^2 \, g^{\mu\nu} \frac{1}{q_1^2} \frac{1}{2p\cdot q_1-q_1^2} \frac{1}{2p\cdot q_1+q_1^2} \frac{ 
(2\epsilon^2-3\epsilon+1) m^2 q_1^2 + 2(2\epsilon-3)m^4 - 2(\epsilon-1) (p\cdot q_1)^2 
}{(2\epsilon-3)m^2} \label{resTIR}
\end{equation}
as one can see the tensor integrals are reduced to the scalar integrals, and it is ready to perform the IBP reduction on those integrals which will be introduced in the next section.

\subsection{Partial fraction decomposition and IBP reduction}

Usually, one gets the scalar loop integrals with propagators that form an overdetermined basis, for example, one of the scalar integrals appearing in (\ref{resTIR}) is as follows:
\begin{equation}
{\cal I} = \int \frac{d^D q_1}{(2\pi)^D} (q_1^2)^{-1} (2p\cdot q_1-q_1^2)^{-1} (2p\cdot q_1+q_1^2)^{-1} \label{apart_eq1}
\end{equation}
if we take the scalar products involving loop momentum ({\it i.e.}, $q_1^2$ and $p\cdot q_1$) as independent vectors, then the three propagators $q_1^2$ and $(2p\cdot q_1 \pm q_1^2)$ are linear dependent. The original {\tt Mathematica} package {\tt Apart} can be used to carry out the partial fraction decomposition automatically on such integrals, these operations have been reimplemented in {\tt HepLib} too, through the function {\tt HepLib::Apart} introduced as follows:
\begin{lstlisting}[language=cpp]
	ex Apart(const ex &expr_in, const lst &loop_ps, const lst & ext_ps);
\end{lstlisting}
where {\tt loop\ttus{}ps} and {\tt ext\ttus{}ps} are the same as those in {\tt TIR} function introduced in last section. One can carry out the partial fraction decomposition on (\ref{apart_eq1}) using the following snip of the code {\tt 7.cpp}:
\begin{lstlisting}[language=cpp]
	Vector p("p"), q1("q1");
    ex expr = 1/SP(q1) * 1/(2*SP(p,q1)-SP(q1)) * 1/(2*SP(p,q1)+SP(q1));
    ex r = Apart(expr,lst{q1},lst{p});
    ex r1 = ApartIR2ex(r); ex r2 = ApartIR2F(r);
\end{lstlisting}
where the returned expression {\tt r} from {\tt Apart} involves the {\it pseudo-function} {\tt ApartIR} (an internal matrix representation) which can be transformed into a normal {\tt ex} expression or {\tt F} function (as in the {\tt FIRE} program) with the help of {\tt ApartIR2ex} and {\tt ApartIR2F} respectively, for example, the results {\tt r1} and {\tt r2} from {\tt ApartIR2ex} and {\tt ApartIR2F} look as follows:
\begin{lstlisting}[language=cpp]
	r1: (-1/2)*q1.q1^(-2)*(q1.q1+2*q1.p)^(-1)+(1/2)*q1.q1^(-2)*(2*q1.p+(-1)*q1.q1)^(-1)
	r2: (-1/2)*F({q1.q1,q1.q1+(-2)*q1.p},{2,1})+(-1/2)*F({q1.q1,q1.q1+2*q1.p},{2,1})
\end{lstlisting}
and in more readable form as follows:
\begin{equation}
-\frac{1}{2}(q_1^2)^{-2}(q_1^2+2p\cdot q_1)^{-1} - \frac{1}{2}(q_1^2)^{-2}(q_1^2-2p\cdot q_1)^{-1} \label{int_eq1}
\end{equation}
where one can see that there are only two independent propagators in each term after the decomposition. 
Furthermore, there is a more general version of {\tt Apart} function introduced as follows: 
\begin{lstlisting}[language=cpp]
	ex Apart(const ex &expr_in, const lst &vars, exmap sign_map={});
\end{lstlisting}
where {\tt vars} refers to a list of independent vectors, and the internal details can be found in \cite{Feng:2012iq}, here one can also get the same result as {\tt r} with the following function call:
\begin{lstlisting}[language=cpp]
	rr = Apart(expr, lst{SP(q1),SP(p,q1)}); // rr is the same as r
\end{lstlisting}

Now it is ready to perform the actual IBP reduction by the functions or classes introduced under the namespace {\tt HepLib::IBP}. The classes {\tt FIRE} and {\tt KIRA} are introduced in the current version, both are derived from the {\tt Base} class:
\begin{lstlisting}[language=cpp]
	class Base {
    public:
        lst Internal; lst External; lst Replacements; lst Propagators;
        lst Integrals; lst Cuts; lst Rules;  lst MIntegrals;
        virtual void Export(); virtual void Run(); virtual void Import();
        void Reduce();
    };
\end{lstlisting}
where the function {\tt Reduce} is just sequential calls to the virtual functions {\tt Export}, {\tt Run} and {\tt Import}.
The classes {\tt FIRE} and {\tt KIRA} have implemented the virtual function {\tt Export} to prepare the input information for calling {\tt FIRE} and {\tt KIRA} programs, the virtual function {\tt Run} to invoke the shell command {\tt FIRE5}/{\tt FIRE6} or {\tt kira} respectively, and the virtual function {\tt Import} to get the reduced results back to {\tt HepLib}.

The basic usage of each class is similar to {\tt FIRE} program by providing the necessary input parameters, including {\tt Internal} (the loop momenta), {\tt External} (the external momenta), {\tt Propagators} (a list of complete and linear-independent propagators), {\tt Replacements} (the scalar product substitution) and {\tt Integrals} (the integrals needed to be reduced), then calling {\tt Reduce} subroutine, {\tt HepLib} will generate the IBP identities automatically with respect to provided {\tt Internal} and {\tt External}, and export the corresponding files which are needed for the {\tt FIRE} or {\tt KIRA} program. After {\tt Reduce}, the members {\tt Rules} and {\tt MIntegrals} will get updated accordingly if everything works fine. 

We exemplify the basic usage of {\tt FIRE} on the first integral in (\ref{int_eq1}), by the following snip of the code {\tt 8.cpp}:
\begin{lstlisting}[language=cpp]
	Symbol q1("q1"), p("p"), m("m");
    FIRE fire; fire.Internal=lst{ q1 }; fire.External=lst{ p };
    fire.Replacements=lst{ p*p == m*m };
    fire.Propagators=lst{ q1*q1, 2*p*q1-q1*q1 };
    fire.Integrals.append(lst{2,1}); // one can add more integrals
	fire.WorkingDir="IBPdir";
    fire.Reduce(); // fire.Rules and fire.MIntegrals got updated
\end{lstlisting}
where the \cpp version of {\tt FIRE} program is used, both {\tt FIRE6} and {\tt FIRE5} are supported, the default one is to use {\tt FIRE6} (to use {\tt FIRE5}, one needs to set the static member {\tt FIRE::Version} to {\tt 5}). Note that the {\tt .start} files are generated by {\tt HepLib} itself, which are different from the original ones generated from the {\tt Mathematica} package {\tt FIRE.m}. After the function call {\tt fire.Reduce()}, one can get the reduction result from the member {\tt fire.Rules} as follows:
\begin{lstlisting}[language=cpp]
	{ F(0,{2,1})==(1/4)*m^(-4)*F(0,{0,1})*((-5)+d)^(-1)*((-2)+d) }
\end{lstlisting}
which represents the replacing rule $\{ F(0,\{2,1\}) \to (d-2)/(d-5) \times F(0,\{0,1\})/(4m^4) \}$, and the resulting master integrals from {\tt fire.MIntegrals}, and there is only one master integral {\tt F(0,\{0,1\})} in this case.
Usually, one will get a number of master integrals that are not independent to each other, the function {\tt IBP::FindRules} is provided to find the relations between those master integrals, the basic idea is almost the same as the one in {\tt FIRE} package, which uses permuted {\tt UF} function (a routine to calculate Symanzik polynomials) to find identical integrals.

All those operations (from partial fraction decomposition to IBP reduction) are wrapped in a single function call {\tt HepLib::ApartIBP} introduced as follows:
\begin{lstlisting}[language=cpp]
	void ApartIBP(int IBPmethod, exvector &io_vec, const lst & loops_exts=lst{});
\end{lstlisting}
with {\tt IBPmethod=1} for using {\tt FIRE} and {\tt 2} for {\tt KIRA}, the function will call {\tt Apart} for partial fraction decomposition, {\tt Reduce} for IBP reduction and {\tt FindRules} to eliminate identical master integrals. Note that the input argument {\tt io\ttus{}vec} will get updated at the end of the function call.

Now, let's come back to the {\tt amp1} (the amplitude for $Q\bar{Q}\to\gamma^*$ at next-to-leading order), we can summarize all the necessary function calls in the following snip of the code {\tt nlo.cpp}:
\begin{lstlisting}[language=cpp]
	auto amp1 = garRead("amps.gar").op(1).op(0);
    Index mu("mu"), nu("nu");
    letSP(p)=m*m; letSP(p,nu)=0;
    lst extps = lst{ p };
    lst loops = lst{ q1 };
    ex proj = SpinProj(IO::In,1,p1,p2,m,m,nu,-1,-3) * ColorProj(-1,-3);    
    amp1 = amp1 * proj; amp1 = amp1.subs(LI(-2)==mu);
    amp1 = MatrixContract(amp1);
    amp1 = form(amp1);
    amp1 = TIR(amp1, loops, extps);
    amp1 = amp1.subs(lst{NF==3,D==4-2*ep});    
    exvector res_vec;
    res_vec.push_back(amp1);
    ApartIBP(1, res_vec, lst{loops, extps});
    ex res = 0;
    for(auto item : res_vec) res += item.subs(lst{d==4-2*ep,D==4-2*ep});
    ex mi = str2ex("F({q1^2+(-2)*p*q1},{1})"); // the master integral
    ex miv = -tgamma(-1+ep)*pow(m,2*(1-ep))*I*pow(Pi,2-ep)*pow(2*Pi,2*ep-4);
	res = res.subs(lst{ mi==miv }); // substitutions here
	res = series_ex(res,ep,0); // taylor expand to ep^0
\end{lstlisting}
where we use the replacement {\tt subs} to substitute the master integral with its explicit expression, {\it i.e.},
\begin{equation}
\mbox{\tt F({q1\^{}2+(-2)*p*q1},{1})} = \int\frac{dq_1^D}{(2\pi)^D} \frac{1}{q_1^2 - 2p\cdot q_1} = -\frac{i\pi^{D/2}}{(2\pi)^D} \Gamma(-1+\epsilon) m^{2(1-\epsilon)}
\end{equation}
and we end with the following result for the amplitude at next-to-leading order (the {\it bare} result, not including the renormalization yet):
\begin{lstlisting}[language=cpp]
	(-1/4)*sqrt(2)*sqrt(3)*ep^(-1)*gs^2*Pi^(-2)*nu.mu+(1/12)*sqrt(2)*sqrt(3)*gs^2*Pi^(-2)*nu.mu*(4+6*log(m)+(-6)*log(2*Pi)+3*Euler+3*log(Pi))
\end{lstlisting}
and in more readable form as follows:
\begin{equation}
\sqrt{6}\frac{g_s^2}{4\pi^2} g^{\mu\nu} \left[ -\frac{1}{\epsilon} + \gamma_E - \ln(4\pi) + \ln m^2 + \frac{4}{3} \right]
\end{equation}
where the coupling constant $g_s$ is still {\it bare}, not {\it renormalized}, and each {\it bare} $g_s$ will be associated with a $\mu^\epsilon$, so the logarithm $\ln m^2$ will be transformed into $\ln(m^2/\mu^2)$ in the final {\it renormalized} result.
 
One can do similar operations on the amplitudes at next-to-next-to-leading order, while one still needs the renormalization to get the final physical result. One can adopt the multiplicative renormalization, {\it e.g.}, in the amplitudes generated from {\tt QGRAF}, the quark mass in each propagator is treated as a {\it bare} mass $m_0$, its relation to the {\it renormalized} mass $m$ is $m_0 = Z_m m$, where the $Z_m$ is the renormalization constant, the {\tt QCD} related renormalization constants up to two-loop order are taken from \cite{Baernreuther:2013caa} and encoded in the namespace {\tt HepLib::QGRAF::RC}. 

Finally, all the related operations for the process $Q\bar{Q}\to\gamma^*$ up to next-to-next-to-leading order (from amplitude/diagram generation, amplitude evaluation, IBP reduction to the substitution of master integrals, as well as the renormalization) can be found at {\tt codes/nnlo.cpp} in the {\tt HepLib} archive, and one can get the following final expression from the output of {\tt nnlo.cpp}:
\begin{equation}
\frac{{\cal A}_{\rm NNLO}}{{\cal A}_{\rm LO}} = 1-\frac{8}{3}\frac{\alpha_s}{\pi} + \frac{\alpha_s^2}{\pi^2}\left[ \frac{35\pi^2}{54}\left(2\ln\frac{m^2}{\mu^2}-\frac{1}{\epsilon}\right)+\frac{50}{9}\ln\frac{m^2}{\mu^2}-\frac{125\zeta(3)}{9}-\frac{511\pi^2}{324}-\frac{23}{54} -\frac{14}{9}\pi^2\ln2 \right] \label{amp_nnlo}
\end{equation}
where we have substituted $C_A=3$, $C_F=4/3$ and $n_f=4$ (4 active light quarks) for simplicity, and the remaining single pole in (\ref{amp_nnlo}) still needs to be factorized/absorbed into the NRQCD long-distance matrix element, furthermore the $\mu$ in the first logarithm corresponds to the factorization scale $\mu_F$, and $\mu$ in the other logarithm refers to the renormalization scale $\mu_R$, so one arrives at the final expression after these manipulations:
\begin{equation}
1-\frac{8}{3}\frac{\alpha_s}{\pi} + \frac{\alpha_s^2}{\pi^2}\left[ \frac{35\pi^2}{27}\ln\frac{m^2}{\mu_F^2}+\frac{50}{9}\ln\frac{m^2}{\mu_R^2}-\frac{125\zeta(3)}{9}-\frac{511\pi^2}{324}-\frac{23}{54} -\frac{14}{9}\pi^2\ln2 \right]
\end{equation}
and one can check the correctness with \cite{Czarnecki:1997vz,Beneke:1997jm}, {\it e.g.}, the $c_2(m_Q/\mu)$ in \cite{Beneke:1997jm} as follows:
\begin{eqnarray}
c_2(m_Q,\mu) &=& C_F^2 c_{\rm2,A} + C_F C_A c_{\rm2,NA} + C_F T_F n_f c_{\rm2,L} + C_F T_F c_{\rm2,H} \nonumber\\
&=& \frac{35}{27} \pi ^2 \ln \frac{m_Q^2}{\mu^2}-\frac{125 \zeta (3)}{9}-\frac{511 \pi ^2}{324}-\frac{23}{54}-\frac{14}{9} \pi ^2 \ln2
\end{eqnarray}
where the coefficients $c_{\rm2,A}$, $c_{\rm2,NA}$, $c_{\rm2,L}$ and $c_{\rm2,H}$ can be found in \cite{Beneke:1997jm}, and note that $\mu_R = m_Q$ is taken explicitly in those expressions.

\subsection{Numerical evaluation by sector decomposition method}

{\it Sector Decomposition} is a general numerical method used to extract the singularity of Feynman integral within dimensional regularization, we refer the reader to \cite{Heinrich:2008si} for a good review on the method of sector decomposition.  

Principally, sector decomposition can only resolve the singularity located in the boundary of the integration domain, to handle the divergence from the internal of the domain, one usually resorts to the integration contour deformation to avoid crossing the internal singularity. A practical programable contour deformation is to use the following transformation for the corresponding $\cal{F}$-term:
\begin{eqnarray}\label{x2z}
	x_i \to z_i =  x_i - i \lambda_i x_i (1-x_i) \frac{\partial\cal F}{\partial x_i}
\end{eqnarray}
where we have adopted the notation as {\tt FIESTA}, one can always find some \{$\lambda_i$\} for the transformation above, such that the imaginary part of ${\cal F}$ is always negative for the whole integral domain, which is consistent with the $-i\epsilon$-prescription in the original Feynman integral. One also notes that the efficiency for the numerical integration is highly dependent on the chosen \{$\lambda_i$\} in some extreme cases, and there is no priori method to select an optimized \{$\lambda_i$\} yet. The public programs {\tt FIESTA} and {\tt SecDec} have already implemented such a transformation.

We present another implementation in \cpp language with the help of {\tt GiNaC}, different from the package {\tt sector\ttus{}decomposition} which also chooses the {\tt GiNaC} to handle the symbolic operations, we implement the sector decomposition by the geometric strategy with the help of {\tt qhull} library (it is almost a {\tt GiNaC} conversion of {\tt STRATEGY\ttus{}KU2} in {\tt FIESTA}), we also develop some new features, including contour deformation as well.

All related classes or functions are introduced under the namespace {\tt HepLib::SD}, the basic usage is very similar to {\tt FIESTA} program by calling the {\tt Evaluate} method of {\tt SecDec} class after providing the necessary information for the integral. To facilitate the parameter input, two structures are introduced in {\tt HepLib}, one is {\tt FeynmanParameter} which is used to wrap the information of loop integration:
 \begin{lstlisting}[language=cpp]
struct FeynmanParameter {
    lst LoopMomenta; lst tLoopMomenta; lst Propagators; lst Exponents;
    exmap lReplacements; exmap tReplacements; exmap nReplacements; ex Prefactor = 1;
};
\end{lstlisting}
and the other is {\tt XIntegrand} which is introduced for the generic parameter integration:
\begin{lstlisting}[language=cpp]
struct XIntegrand {
	lst Functions; lst Exponents; exmap nReplacements; lst Deltas; bool isAsy = false;
};
\end{lstlisting}

The basic usage involving sector decomposition method can be illustrated by numerically evaluating the integral ${\cal I}_1$ in dimensional regularization with dimension $D=4-2\varepsilon$:
\begin{eqnarray}
{\cal I}_1 = \int \frac{d^Dk\, d^Dr\, d^Dq}{(i\pi^{2-\varepsilon})^3 \Gamma^3(1-\varepsilon)} \frac{1}{(-k^2)^{1+3\varepsilon}} \frac{1}{-(k+p_1+p_2)^2} \frac{1}{-(-k+r)^2} \frac{1}{-(p_1+r)} \frac{1}{-(k-q)^2} \frac{1}{-(p_1+q)^2}
\end{eqnarray}
where $p_1^2=p_2^2=0$ and $s=2p_1\cdot p_2=-1$, using the following snip of the code {\tt 9.cpp}: 
\begin{lstlisting}[language=cpp]
    Symbol k("k"),r("r"),q("q"),p1("p1"),p2("p2"),s("s"); 
    FeynmanParameter fp;
    fp.LoopMomenta = lst{k,r,q};
    fp.Propagators= lst { -pow(k,2),-pow(k+p1+p2,2),-pow(-k+r,2),-pow(p1+r,2),
		-pow(k-q,2),-pow(p1+q,2)};
    fp.Exponents = lst{1+3*ep,1,1,1,1,1};
    fp.lReplacements[p1*p1] = 0;
    fp.lReplacements[p2*p2] = 0;
    fp.lReplacements[p2*p1] = s/2;
    fp.lReplacements[s] = -1;
    fp.Prefactor = pow(I*pow(Pi,2-ep),-3) * pow(tgamma(1-ep),3); 
    SecDec work; work.epN = 0; Verbose = 2; work.Evaluate(fp);
    hout << work.VEResult() << endl; 
\end{lstlisting}
where {\tt Verbose=2} is used for some detailed output, and {\tt epN=0} refers to evaluating the integral to ${\cal O}(\varepsilon^0)$, the {\tt hout} statement will result in the output as follows:
\begin{lstlisting}[language=cpp]
  1.2532095(6)E2+ep^(-3)*1.666666667(4)E-1+ep^(-2)*1.83333333(6)+ep^(-1)*1.812319(2)E1
\end{lstlisting}
which corresponds to 
\begin{eqnarray}
{\cal I}_1 = 125.32095(6)+\varepsilon^{-3}\times0.1666666667(4)+\varepsilon^{-2}\times1.83333333(6)+\varepsilon^{-1}\times18.12319(2) + {\cal O}(\varepsilon)
\end{eqnarray}

One can see that all operations are encoded in the function call to {\tt Evaluate} method of {\tt SecDec}, which actually contains the following basic procedures while using sector decomposition method:
\begin{lstlisting}[language=cpp]
	1. Initialize(fp);		2. Scalelesses();		3. KillPowers();		
	4. RemoveDeltas();		5. KillPowers();		6. SDPrepares();
	7. EpsEpExpands();		8. CIPrepares();		9. Contours();		10.Integrates();
\end{lstlisting}
They are the member functions of class {\tt SecDec} parts of which are listed below:
\begin{lstlisting}[language=cpp]
	class SecDec {
	public:
		int epN = 0;  int epsN = 0; bool CheckEnd = false;
		vector<ex> FunExp;  vector<ex> Integrands;  vector<ex> expResult;
		SecDecBase *SecDec = NULL;  IntegratorBase *Integrator = NULL;  
		MinimizeBase *Minimizer = NULL;
		ex ResultError; 
		void Initialize(FeynmanParameter fpi);  void Initialize(XIntegrand xint);
        void Normalizes();  void Scalelesses();
        void SDPrepares();  void EpsEpExpands();  void RemoveDeltas();
        void CIPrepares(const string & key = "");
        void Contours(const string & key = "", const string & pkey = "");
        void Integrates(const string & key="", const string & pkey="", int kid=0);
        void Evaluate(FeynmanParameter fpi, const string & key = "");
        void Evaluate(XIntegrand xint, const string & key = "");
        void Evaluate(vector<ex> FunExp, const string & key = "");
        void MB();  void XEnd();  void ChengWu();
	}
\end{lstlisting}
The detailed descriptions of the members or functions in {\tt SecDec}, we refer the reader to the document~\cite{HepLib_document}.

The key operations involving sector decomposition methods are sequential calls from {\tt SDPrepares} to {\tt Integrates}: {\tt SDPrepares} is used to perform the sector decomposition currently with the geometric strategy, {\tt EpsEpExpands} is used to expand the integrand around {\tt eps=0} and {\tt ep=0} ({\tt eps} is another infinitesimal regulator similar to the dimensional regulator {\tt ep}, and note that the expansion around {\tt eps=0} will be performed before the expansion of {\tt ep}), {\tt CIPrepares} is used to generate \cpp source code for the integrand, the \cpp code will be compiled to a shared {\tt .so} library, {\tt Contours} is used to perform contour deformation if the ${\cal F}$-term has negative terms, {\tt Integrates} is used to perform the actually numerical integrations, with the help of {\tt dl} library, the integrands are dynamically loaded from the compiled {\tt .so} files generated in {\tt CIPrepares}.

{\tt HepLib} currently supports two numerical integrators {\tt CUBA} and {\tt HCubature}~\cite{cubature_website}, a tiny modified version\footnote{A {\tt PrintHooker} has been added to the original version of {\tt HCubature} to retrieve the intermediate integral result and estimated error.} of {\tt HCubature} is shipped along with the library, the interface to each integrator is derived from the base class {\tt IntegratorBase}~\cite{HepLib_document} which wraps a few basic parameters to control the numerical integration. Since the quadruple precision is adopted in the base class {\tt IntegratorBase}, the library {\tt libcubaq} is actually used. Although the default precision of the numerical integrator is quadruple, one can still use other numerical precision to evaluate the integrand at a specific input point and cast the computed values to quadruple precision at the end. There are three internal types of numeric precision in the {\tt HepLib}: {\tt long double}, {\tt quadruple} and {\tt arbitrary} precision by {\tt MPFR}, one of the precisions will be chosen according to the input arguments supplied to the integrand function call, the controlling parameters can be found in the {\tt IntegratorBase}.

For negative ${\cal F}$-terms, {\tt HepLib} adopts contour deformation method to avoid the singularity inside the integration domain, the $\lambda$-tuning is performed dynamically during numerical integration, this is achieved by exporting related ${\cal F}$-terms and their derivatives to the generated \cpp source, and the variable substitution defined in (\ref{x2z}) actually occurs at evaluation running time, instead of the static compilation, generally it will results in a smaller size of source code and a faster running time by avoiding evaluating the same variable substitution multiple times.

In addition, we want to address some other methods than contour deformation, including {\tt WickRotation} and {\tt ChengWu} method, by using the following integral ${\cal I}_2$:
\begin{eqnarray}
{\cal I}_2 = \int_0^1 dx_1 \int_0^1 dx_2 \int_0^1 dx_4\; x_1^{-\frac{5}{2}+3\varepsilon} x_2^{-\frac{3}{2}+\varepsilon} \left( -4 x_4^2-12x_1x_4+x_1x_2 \right)^{\frac{1}{2}-2\varepsilon} \delta(x_1+x_2+x_4-1) 
\end{eqnarray}

The {\tt WickRotation} method is illustrated by the following snip of the code {\tt wr.cpp}:
\begin{lstlisting}[language=cpp]
	ex wra = WRA(Pi/5);
    XIntegrand xint;
    xint.Functions = lst{1,-4*x(4)*x(4)-12*x(1)*x(4)+x(1)*x(2),x(1),x(2)};
    xint.Exponents = lst{1,1/ex(2)-2*ep,3*ep-5/ex(2),ep-3/ex(2)};
    xint.Deltas = lst{lst{x(1),x(2),x(4)}};
    xint.Functions = ex_to<lst>(subs(xint.Functions,{x(4)==x(4)*exp(I*wra)}));
    xint.Functions.let_op(0) = exp(I*wra); // Jacobi 
    SecDec work;    
    work.Initialize(xint);
	... ...
\end{lstlisting}
where we have used {\tt XIntegrand} for the generic parameter integration in sector decomposition method, note that {\tt HepLib} adopts the notation used in {\tt FIESTA}: the ${\cal F}$-term is always supplied at {\tt 2nd} position of {\tt Functions} and compatible with $-i\epsilon$ convention.
In {\tt WickRoation} evaluation of the integral ${\cal I}_2$, we can first pull out $x_4$ from the $\delta$-function with the help of Cheng-Wu theorem~\cite{cheng_wu,smirnov_book}, then make the transformation $x_4 \to x_4 e^{i\pi/5}$ with $\pi/5$ as the rotation angle, finally add $x_4$ back to the $\delta$-function, it is ready to see the imaginary of the ${\cal F}$-term is compatible with $-i\epsilon$ convention when the rotation angle changes from $0$ to $\pi/5$.
The calculated numerical value of ${\cal I}_2$ up to ${\cal O}(\varepsilon)$ (produced by the {\tt codes/wr.cpp}) is as follows:
\begin{eqnarray}
{\cal I}_2 = -\frac{1.5}{\varepsilon} + (3.29583687(5)-4.71238898(5)i) +(14.716159(1)+10.354177(1)i)\varepsilon + {\cal O}(\varepsilon^2)
\end{eqnarray}

The integral ${\cal I}_2$ can also be evaluated by the {\tt ChengWu} method by the snip code {\tt cw.cpp}
\begin{lstlisting}[language=cpp]
    XIntegrand xint;
    xint.Functions = lst{1,-4*x(4)*x(4)-12*x(1)*x(4)+x(1)*x(2),x(1),x(2)};
    xint.Exponents = lst{1,1/ex(2)-2*ep,3*ep-5/ex(2),ep-3/ex(2)};
    xint.Deltas = lst{lst{x(1),x(2),x(4)}};
    SecDec work;    
    work.Initialize(xint);
	work.ChengWu(); // call ChengWu method
	... ...
\end{lstlisting}
note the function call to {\tt ChengWu} method, by Cheng-Wu theorem, $x_1$ and $x_2$ can be pulled out of the $\delta$-function, sequential transformations are then performed automatically in {\tt HepLib}: (1) $x_1 \to x_1^\prime/3$, (2) $x_2 \to 4x_4x_2^\prime/x_1=12x_4x_2^\prime/x_1^\prime$, with Jacobi terms properly added, the corresponding transformed ${\cal F}$-term now becomes $(x_2^\prime-x_1^\prime-x_4)$ (generally, the transformed ${\cal F}$-term will be $(x_1+\cdots+x_m-x_{m+1}-\cdots-x_n)$), again the transformation $x_2^\prime \to y_2 (x_1^\prime+x_4)/y_1$ is made to result in the final ${\cal F}$-term $(y_1-y_2)$, note that the positive coefficient of ${\cal F}$-term is always factorized out. Both $y_1$ and $y_1$ are integrated from $0$ to $\infty$, so we can divide the domain into $y_1\ge y_2$ and $y_1\le y_2$, make transformation $y_1\to y_1+y_2$ and $y_2\to y_1+y_2$ for those two regions respectively, the resulted ${\cal F}$-term will be proportional to $y_1$ and $(-y_2)$, and the negative one can be readily changed to positive with an overall constant factor using $(-y_2-i\epsilon)^{a+b\varepsilon} = e^{-i\pi(a+b\varepsilon)} y_2^{a+b\varepsilon}$, finally $y_1$ or $y_2$ is changed back to $x_1$ or $x_2$ respectively, which will be added back to $\delta$-function with the help of Cheng-Wu theorem. All those internal operations are actually performed by the functions from a separate class {\tt ChengWu}~\cite{HepLib_document}.
Furthermore, the integral ${\cal I}_2$ can also be cross-checked by contour deformation by using the code {\tt codes/ct.cpp} in {\tt HepLib} archive.

Note that, when the integrand does not encounter any singularity in the whole domain with the help of contour deformation or the {\tt WickRoation} method, the numerical integration itself does not pose any problem and always results in a finite result, however we need pay attention to the branch cut of multiple-valued functions (including the {\tt log} or {\tt power}). {\tt HepLib} treats the branch cut in a simple way, taking a {\tt log} term $\ln(g(z_i(\vec{\lambda}))$ in the method of contour deformation as an example, we multiple $\vec{\lambda}$ by a scaling parameter $s$, when the parameter $s$ varies from $0$ to $1$, we can check whether the argument of {\tt log} function $g(z_i(\vec{\lambda}))$ goes through the branch cut or not. Practically, we let $s$ take $n$ values $s_j = j/n$ ($0<j\le n$), then compare the corresponding $g(z_i(s_k\vec{\lambda}))$ and $g(z_i(s_{k+1}\vec{\lambda}))$, if the line segment between the two points goes through the branch cut, a $(2\,i\,\pi)$ is added or subtracted accordingly.

Finally, we want to demonstrate the automatic expansion of loop/parameter integrals with respect to a small parameter by combining the sector decompositions and Mellin–Barnes representations (initially suggested in \cite{Pilipp:2008ef}), by using the example from \cite{Smirnov:2009pb} in the following snip of the code {\tt exp.cpp}:
\begin{lstlisting}[language=cpp]
	auto t = vs; // the global Symbol vs 
	FeynmanParameter fp;
    fp.LoopMomenta = lst{ k };
    fp.Propagators = lst{ -pow(k,2),-pow(k+p1,2),-pow(k+p1+p2,2),-pow(k+p1+p2+p4,2) };
    fp.Exponents = lst{ 1, 1, 1, 1 };
    fp.lReplacements[p2*p4] = -t/2;
    fp.lReplacements[p1*p4] = (s+t)/2;
	... ...
	SecDec work;
	work.Evaluate(fp);
\end{lstlisting}
the only difference from the unexpanded case is to use the global {\tt Symbol} {\tt vs} to indicate the small parameter, {\tt SecDec} class will handle the expansion automatically and result in the following expansion:
\begin{eqnarray}
\int\frac{d^{4-2\varepsilon}k}{i\pi^{2-\varepsilon}e^{-\varepsilon\gamma_E}} \frac{1}{k^2} \frac{1}{(k+p_1)^2} \frac{1}{(k+p_1+p_2)^2} \frac{1}{(k+p_1+p_2+p_4)^2} = \frac{4}{\varepsilon^2 t}-\frac{2\ln t}{\varepsilon t}-\frac{13.159472535(0)}{t}+{\cal O}(\varepsilon,t) 
\end{eqnarray}
with the kinematics $p_1^2=p_2^2=p_4^2=0$, $p_1\cdot p_2=-s/2$, $p_2\cdot p_4=-t/2$, $p_1\cdot p_4=(s+t)/2$ and $s=1$.

\section{Installation and Usage}
Before we go into the details of the installation, one can use the script {\tt install.sh} or {\tt makefile} to install the required external libraries and programs automatically, it will install {\tt HepLib} as well, by invoking one of the following commands in the terminal:
\begin{itemize}
\item using {\tt install.sh}
\begin{lstlisting}[language=cpp]
	wget https://heplib.github.io/install.sh
	chmod +x install.sh
	INSTALL_PATH=<INSTALL PATH> jn=<jn> ./install.sh
\end{lstlisting}
\item using {\tt makefile}
\begin{lstlisting}[language=cpp]
	wget https://heplib.github.io/makefile
	make INSTALL_PATH=<INSTALL PATH> jn=<jn>
\end{lstlisting}
\end{itemize}
where the {\tt <INSTALL PATH>} is the path for the libraries to be installed to, and {\tt <jn>} is the number of jobs used in invoking {\tt make -j \$jn} in the script.

Since {\tt GiNaC} or {\tt HepLib} will get updated from time to time, one can use the {\tt update} script (located in the directory {\tt <INSTALL PATH>/bin/}) to only update {\tt GiNaC} or {\tt HepLib} to their latest versions:
\begin{lstlisting}[language=cpp]
	jn=<jn> <INSTALL PATH>/bin/update ginac  # update both GiNaC and HepLib
	jn=<jn> <INSTALL PATH>/bin/update heplib # only update HepLib
\end{lstlisting}

\subsection{Prerequisites}
Since {\tt HepLib} uses several external routines or libraries, one needs to install these required libraries before the installation of {\tt HepLib}, one also needs {\tt cmake} and {\tt GNU} compiler system for the compilation as well. Those prerequisites can be divided into two categories: {\it{}external libraries} and {\it external programs}, only {\it external libraries} are required for the compilation or installation of {\tt HepLib}, while to run the user program properly which uses {\tt HepLib}, it's still required for the {\it external programs} to be found in the environment variable {\tt PATH}. Note that there is no need to adjust the variable {\tt PATH} if one installs {\tt HepLib} using the install script {\tt install.sh} or {\tt makefile}.

Before the installation, one needs to provide the installation destination through the environment variable {\tt \$INSTALL\ttus{}PATH} by typing the commands in the terminal:
\begin{lstlisting}[language=cpp]
	  export INSTALL_PATH="<INSTALL PATH>"
\end{lstlisting}

\begin{itemize}
\item {\it External libraries:}
\begin{enumerate}
\item {\tt GMP} is required by {\tt MPFR} and {\tt GiNaC}. The typical instructions to download and install {\tt GMP} (v6.2.1) is as follows:
\begin{lstlisting}[language=cpp]
curl -L -O https://gmplib.org/download/gmp/gmp-6.2.1.tar.bz2
tar jxf gmp-6.2.0.tar.bz2;cd gmp-6.2.1
./configure --prefix=$INSTALL_PATH 
make -j 16 && make install
\end{lstlisting}

\item {\tt MPFR} is used to handle the multiple precision in the numerical integration when large number cancelation occurs, since we choose quadruple precision as the default float precision type in the numerical integrator, {\tt MPFR} needs to be compiled with the option {\tt --enable-float128}. Note that the quadruple precision type {\tt \ttus\ttus{}float128} has been changed to {\tt \ttus{}Float128} since {\tt MPFR 4.1.0}, so we prefer the version {\tt MPFR 4.0.2} for the moment, furthermore the {\tt MPFR C++}~\cite{mpfr_cpp_website} wrapper is included in {\tt HepLib} archive. The typical instructions to download and install {\tt MPFR} (v4.0.2) is as follows:
\begin{lstlisting}[language=cpp]
curl -L -O https://heplib.github.io/download/mpfr-4.0.2.tar.gz
tar zxf mpfr-4.0.2.tar.gz;cd mpfr-4.0.2
./configure --prefix=$INSTALL_PATH --enable-float128 --enable-thread-safe 
make -j 16 && make install
\end{lstlisting}

\item {\tt CLN} is required by {\tt GiNaC}. The typical instructions to download and install {\tt CLN} (v1.3.6) is as follows:
\begin{lstlisting}[language=cpp]
curl -L -O https://www.ginac.de/CLN/cln-1.3.6.tar.bz2
tar jxf cln-1.3.6.tar.bz2;cd cln-1.3.6
./configure --prefix=$INSTALL_PATH --with-gmp=$INSTALL_PATH
make -j 16 && make install
\end{lstlisting}

\item {\tt GiNaC} is actually the underlying language of {\tt HepLib}, all symbolic functions are built on top of it. The typical instructions to download and install {\tt GiNaC} (v1.8.0) is as follows:
\begin{lstlisting}[language=cpp]
curl -L -O https://www.ginac.de/ginac-1.8.0.tar.bz2
tar jxf ginac-1.8.0.tar.bz2;cd ginac-1.8.0
./configure --prefix=$INSTALL_PATH PKG_CONFIG_PATH=$INSTALL_PATH/lib/pkgconfig
make -j 16 && make install
\end{lstlisting}

\item {\tt QHull} is used in the sector decomposition when using the geometric strategy. The typical instructions to download and install {\tt QHull} (v2020.2) is as follows:
\begin{lstlisting}[language=cpp]
curl -L -O http(-@{}@-)://www.qhull.org/download/qhull-2020.2.zip
unzip -q qhull-2020.2.zip;cd qhull-2020.2
cp Makefile Makefile.bak
cat Makefile.bak | sed "s/\/usr\/local/\$\$INSTALL_PATH/g" > Makefile
make && make install
\end{lstlisting}

\item {\tt MinUit2} is used to find the numerical minimum of a function. The typical instructions to download and install {\tt MinUit2} (v5.34.14) is as follows:
\begin{lstlisting}[language=cpp]
curl -L -O http(-@{}@-)://project-mathlibs.web.cern.ch/project-mathlibs/sw/5_34_14/Minuit2/Minuit2-5.34.14.tar.gz
tar zxf Minuit2-5.34.14.tar.gz;cd Minuit2-5.34.14
./configure --prefix=$INSTALL_PATH
make -j 16 && make install
\end{lstlisting}

\item {\tt CUBA} is one of the numerical integrators with many different numerical integration routines, note that {\tt libcubaq} is actually used by providing the option {\tt --with-real=16} to the {\tt configure} script. The typical instructions to download and install {\tt CUBA} (v4.2) is as follows:
\begin{lstlisting}[language=cpp]
curl -L -O http(-@{}@-)://www.feynarts.de/cuba/Cuba-4.2.tar.gz
tar zxf Cuba-4.2.tar.gz;cd Cuba-4.2
./configure --prefix=$INSTALL_PATH --with-real=16 CFLAGS="-fPIC -fcommon"
make && make install
\end{lstlisting}

\end{enumerate}
\item {\it External programs:}
\begin{enumerate}
\item {\tt Fermat} is used for high-performance operations on matrix or rational polynomial. The typical installation instructions for {\tt Linux} are as follows:
\begin{lstlisting}[language=cpp]
curl -L -O http(-@{}@-)://home.bway.net/lewis/fermat64/ferl6.tar.gz
tar zxf ferl6.tar.gz;mv ferl6 $INSTALL_PATH;
cd $INSTALL_PATH/bin;ln -s -f ../ferl6/fer64 .
\end{lstlisting}

\item {\tt FORM} is used for Dirac/Color matrix trace, Lorentz index contraction and so on. The typical installation instructions for {\tt Linux} are as follows:
\begin{lstlisting}[language=cpp]
curl -L -O https://github.com/vermaseren/form/releases/download/v4.2.1/form-4.2.1-x86_64-linux.tar.gz
tar zxf form-4.2.1-x86_64-linux.tar.gz
cp -rf form-4.2.1-x86_64-linux/form $INSTALL_PATH/bin/
cp -rf form-4.2.1-x86_64-linux/tform $INSTALL_PATH/bin/
\end{lstlisting}

\item {\tt FIRE} is required for IBP reduction using {\tt FIRE} class. The typical installation instructions for {\tt Linux} are as follows:
\begin{lstlisting}[language=cpp]
git clone https://bitbucket.org/feynmanIntegrals/fire.git
mv fire/FIRE6 $INSTALL_PATH/FIRE6
cd $INSTALL_PATH/FIRE6
./configure && make -j 16 && make
\end{lstlisting}

\item {\tt KIRA} is used for IBP reduction using {\tt KIRA} class. The typical installation instructions for {\tt Linux} are as follows:
\begin{lstlisting}[language=cpp]
curl -L -o kira https://kira.hepforge.org/downloads?f=binaries/kira-2.0
chmod +x kira;mv -f kira $INSTALL_PATH/bin/kira
\end{lstlisting}

\end{enumerate}
\end{itemize}

\subsection{Installation of {\tt HepLib}}
As the required external libraries have been installed, the most recent version of {\tt HepLib} is available for download from \burl{https://heplib.github.io/HepLib.tar.gz} as a compressed archive. One can uncompress the archive and change the current directory into the extracted directory {\tt HepLib/src} by typing the commands in the terminal:
\begin{lstlisting}[language=cpp]
	wget https://heplib.github.io/HepLib.tar.gz
	tar zxfv HepLib.tar.gz;cd HepLib/src
\end{lstlisting}
and create a directory {\tt build} for {\tt cmake} to build the library as follows:
\begin{lstlisting}[language=cpp]
	mkdir build && cd build
	cmake -DCMAKE_INSTALL_PREFIX=<INSTALL PATH> ..
	make -j 4 && make install
\end{lstlisting}
where the standard {\tt cmake} variable {\tt CMAKE\ttus{}INSTALL\ttus{}PREFIX} refers to the directory to which {\tt HepLib} will be installed, {\it i.e.}, the library {\tt libHepLib.so} (the file name may be system dependent) will be installed to {\tt <INSTALL PATH>/lib}, the related \cpp header files will be installed to {\tt <INSTALL PATH>/include}, and the binary programs, including {\tt heplib++}, {\tt garview}, {\it etc.}, will be installed to {\tt <INSTALL PATH>/bin}.

If {\tt GiNaC} or other dependent external library is not installed to {\tt CMAKE\ttus{}INSTALL\ttus{}PREFIX}, one needs to specify the locations by supplying the variables {\tt INC\ttus{}PATH} and {\tt LIB\ttus{}PATH} in the {\tt cmake} arguments as follows:
\begin{lstlisting}[language=cpp]
  cmake -DCMAKE_INSTALL_PREFIX=path -DINC_PATH="inc1;inc2" -DLIB_PATH="lib1;lib2" ..
\end{lstlisting}

The most recent version of the detailed document is available on the website \burl{https://heplib.github.io/doxygen/}, and to view the html document on the local machine, one can run the shell command {\tt doxygen} in the directory {\tt HepLib/src/doc} and open {\tt html/index.html} in a browser.

It is not required but will be convenient for later use to adjust the related environment variable by appending the following command to one's {\tt .bashrc}:
\begin{lstlisting}[language=cpp]
	export PATH=<INSTALL PATH>/bin:$PATH
\end{lstlisting}

\subsection{Basic usage within \cpp}
The basic usage of {\tt HepLib} is similar to other \cpp library by including the proper header files in the \cpp source code, compiling the program and linking with {\tt HepLib} and other necessary libraries. The following code shows how to perform the $D$-dimensional trace on a Dirac-$\gamma$ chain ${\rm Tr}[p\!\!\!/_1 \gamma^\mu (p\!\!\!/_2+m) \gamma_\mu]$ and print the result at the end:
\begin{lstlisting}[language=cpp]
	#include "HepLib.h"
	using namespace HepLib;
	int main(int argc, char** argv) {
	    Index mu("mu"), nu("nu");
	    Vector p1("p1"), p2("p2");
	    Symbol m("m");
	    ex gline = GAS(p1)*GAS(mu)*(GAS(p2)+m*GAS(1))*GAS(mu); 
	    ex trace = form(TR(gline));
	    hout << trace << endl;
	    return 0;
	}
\end{lstlisting}
Assuming the file is named {\tt trace.cpp}\footnote{{\tt trace.cpp} can be found in the directory {\tt codes} in {\tt HepLib} archive}, one can compile and run it as follows:
\begin{lstlisting}[language=cpp]
	$ <INSTALL PATH>/bin/heplib++ -o trace trace.cpp
	$ ./trace 
	8*p2.p1+(-4)*D*p2.p1
\end{lstlisting}
where {\tt heplib++} is a shell script installed to {\tt <INSTALL PATH>/bin}, {\tt heplib++} encodes the {\tt g++} flags including {\tt -I}, {\tt -L} and {\tt -l} flags. 
One can also compile the program with {\tt pkg-config} as follows:
\begin{lstlisting}[language=cpp]
	export PKG_CONFIG_PATH=<INSTALL PATH>/lib/pkgconfig:$PKG_CONFIG_PATH
	g++ $(pkg-config --cflags --libs HepLib) -o trace trace.cpp
\end{lstlisting}
or provide the {\tt g++} flags explicitly as follows:
\begin{lstlisting}[language=cpp]
g++ -I <INSTALL PATH>/include -L <INSTALL PATH>/lib -Wl,-rpath,<INSTALL PATH>/lib -lHepLib -lginac -o trace trace.cpp
\end{lstlisting}

\subsection{Examples}
To facilitate the usage of {\tt HepLib} library, a list of examples are also shipped along with {\tt HepLib.tar.gz},  here we provide a few descriptions for helping to use the examples for the current version.

\begin{itemize}

\item The files in {\tt examples/Basic} provide some basic usages of the functions extended to {\tt GiNaC} itself, including the interfaces to {\tt Fermat} and {\tt FORM} programs.

\item The files in {\tt examples/Hep} demonstrate the basic objects in High Energy Physics, including a complete example using {\tt qgraf}, usage of {\tt form} to evaluate Dirac-$\gamma$ or {\tt SUNT} traces, {\tt TIR} for tensor index reduction, IBP reductions with {\tt FIRE} and so on.

\item The files in {\tt examples/SD} illustrate basic usage of sector decomposition method, including numerical evaluation of loop integrals or general parametric integration using {\tt SecDec} class. 
\end{itemize}

More examples to be added in the future will be available at \burl{https://heplib.github.io/example.html}.

\section{Summary and outlook}
A \cpp library {\tt HepLib} is presented for computations in High Energy Physics, it uses {\tt GiNaC} for the symbolic operations. The library tries to provide an integrated \cpp interface to several well-established programs, including {\tt qgraf} for generating Feynman diagrams and amplitudes, {\tt Fermat} for fast simplifications on rational polynomials, {\tt FORM} for highly efficient calculations of Dirac-$\gamma$ and Color matrix. The geometric sector decomposition method has also been implemented in the native {\tt GiNaC} language with the help of {\tt qhull} library with many new features.
In the future version, we are going to introduce the namespace {\tt HepLib::DE} for the related operations using DE method, the namespace {\tt HepLib::MB} for the related operations using MB method, other IBP reduction programs (like {\tt Reduze}) to {\tt HepLib::IBP}, interfaces to other programming languages (including {\tt Mathematica}, {\tt Python}, {\it etc.}) and many other features.

\section*{Acknowledgment}
The authors are grateful to the developers of {\tt GiNaC} for answering questions and providing supports on the {\tt GiNaC} mailing list~\cite{GiNaC_list_archives}.
This work is supported by the National Natural Science Foundation of China under Grants No. 11875318, and by the Yue Qi Young Scholar Project in China University of Mining and Technology (Beijing).

\end{document}